\pdfoutput=1
\documentclass[structabstract]{aa}  

\usepackage{longtable,lscape}
\usepackage{graphicx}
\usepackage{natbib}
\usepackage{txfonts}
\usepackage{multirow}
\begin{document}
   \title{Structure of the hot molecular core G10.47+0.03}

   \author{R. Rolffs\inst{1,2}
          \and
          P. Schilke\inst{2}
          \and
          Q. Zhang\inst{3}
          \and
           L. Zapata\inst{1,4}
          }

   \institute{Max-Planck-Institut f\"ur Radioastronomie,
              Auf dem H\"ugel 69, 53121 Bonn, Germany\\
              \email{rrolffs@mpifr.de}
         \and
             I. Physikalisches Institut, Universit\"at zu K\"oln,
  Z\"ulpicher Stra\ss e 77, 50937 K\"oln, Germany\\
             \email{schilke@ph1.uni-koeln.de}
 	\and
	     Harvard-Smithsonian Center for Astrophysics, 60 Garden Street, Cambridge, MA 02138, USA
	\and
	     Centro de Radioastronom\'ia y Astrof\'isica, UNAM, Apdo. Postal 3-72 (Xangari), 58089 Morelia, Michoac\'an, Mexico
         }

   \date{Received ?}

  \abstract
% Context
{The physical structure of hot molecular cores, where forming massive stars 
have heated up dense dust and gas, but have not yet ionized the molecules, 
poses a prominent challenge in the research of high-mass star formation and 
astrochemistry. }
% Aims
{We aim at constraining the spatial distribution of density, temperature, 
velocity field, and chemical abundances in the hot molecular core 
G10.47+0.03.  }
% Methods
{With the Submillimeter Array (SMA), we obtained high spatial and spectral 
resolution of a multitude of molecular lines at different frequencies, 
including at 690 GHz. At 345 GHz, our beam size is $0.3''$, corresponding to 
3000 AU. We analyze the data using the three-dimensional dust and line 
radiative transfer code RADMC-3D for vibrationally excited HCN, and myXCLASS for line identification.}
% Results
{We find hundreds of molecular lines from complex molecules and high 
excitations. Even vibrationally excited HC$^{15}$N at 690 GHz is detected. The HCN abundance at high temperatures is very high, on the order of 10$^{-5}$ relative to H$_2$.
Absorption against the dust continuum occurs in twelve transitions, whose 
shape implies an outflow along the line-of-sight. Outside the continuum peak, the line shapes are indicative of infall. Dust continuum and 
molecular line emission are resolved at 345/355 GHz, revealing central 
flattening and rapid radial falloff of the density outwards of 10$^4$ AU, best reproduced by a Plummer radial profile of the density.
No fragmentation is detected, but modeling of the line shapes of vibrationally excited HCN suggests the density to be clumpy.
}
% Conclusions
{We conclude that G10.47+0.03 is characterized by beginning of
feedback from massive stars, while infall is ongoing.  Large gas masses (hundreds of M$_\odot$) are heated to high temperatures above 300 K, aided by diffusion of radiation  in a high-column-density environment. The increased thermal, radiative, turbulent, and wind-driven pressure drives expansion in the central region and is likely responsible for the central flattening of the density.}

   \keywords{ISM: molecules  --
        ISM: structure --
      ISM: clouds --
    Stars: formation}

\maketitle

%
%________________________________________________________________

\section{Introduction}

Massive stars and star clusters are born deeply embedded in molecular clouds
\citep[for a review, see][]{ZinneckerYorke07}. When cores have sufficiently
contracted to form massive stars, the dense dust and gas is heated by these
stars. The ice mantles around dust grains evaporate, and a rich plethora of
molecular lines can be observed, along with strong dust emission \citep{Kurtz00,Cesaroni05}. Often, these hot molecular cores are associated with ultracompact or hypercompact H{\sc ii} regions, which are ionized by newly formed massive stars \citep{Hoare07}. The physical and chemical structure of hot molecular cores is of great importance for the study of high-mass star formation and of astrochemistry. Investigations of the structure are hampered, though, by the compactness of the sources, by their scarcity and large distance, and by the large foreground column density, which only long-wavelength radiation can pass.

What is needed, hence, is high angular resolution at (sub)millimeter
wavelengths, combined with high spectral resolution of many molecular lines,
which contain all the information about chemistry, velocity field, and
temperature. With the Submillimeter Array (SMA) in Hawaii, we observed the massive hot molecular core G10.47+0.03 at around 200, 345, and 690 GHz, yielding a best resolution of
$0.3''$ and  350  identified molecular lines. Due to its large dust mass and
high temperatures, this hot core is among the strongest sources at submillimeter
wavelengths, although it is located at a distance of 10.6 kpc
\citep{Pandian08}. It has an estimated luminosity of $7\times 10^5$ L$_\odot$
\citep{Cesaroni10} and displays exceptionally many lines from highly excited
molecules \citep[e.g. HC$_3$N, ][]{Wyrowski99}.

\begin{figure}
  \centering
  \includegraphics[bb=60        417        219        735,angle=0,width=0.4\textwidth]{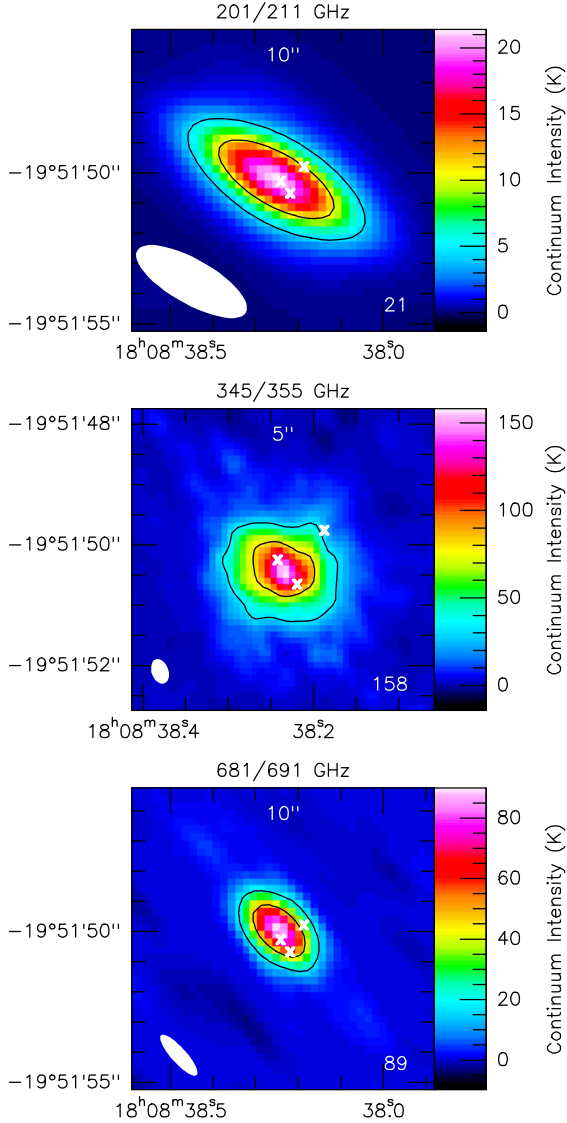} 
 \caption{Continuum maps of G10.47+0.03 observed with the SMA at 201/211 GHz (top panel), 345/355 GHz (central), and 681/691 GHz (bottom). The map size is either 5 or 10$''$. Contours mark 20 and 50\% of the peak flux, which is given in the lower right  (in K). The beam is depicted in the lower left. The white crosses denote the H{\sc ii} regions B1, B2, and A \citep[from left to right]{Cesaroni10}. }
  \label{fig:cont}
\end{figure}

\section{Observations and data reduction}\label{sec:obs}

\begin{table*}
\caption[]{Observational summary}
\label{tab:obs_dates}
\begin{tabular}{l c cc c c c}
\hline\hline
Date  & frequency & array configuration & baseline\tablefootmark{a} & no. antennas  & integration time & gain calibrator\\
~ &       (GHz) & ~ &  (m) & ~ & (min) & \\
\hline
12 Aug 2007 & 345/355 & extended &  205              & 7 & 281 &  1733-130    \\
10 May 2008 & 681/691 + 201/211 & compact north & 125& 5 & 270 & Callisto     \\
6 Sep 2008 & 681/691 + 345/355 &  subcompact & 70   & 6 & 82 &  Callisto  \\
7 Sep 2008 & 681/691 + 345/355 & subcompact &  70   & 7 & 93 &  Callisto, Ganymede \\
4 July 2009 & 345/355 &  very extended &     516    & 8 & 330 &  1733-130   \\
\hline
\end{tabular}
\tablefoot{
\tablefoottext{a}{Maximum projected baseline}
}
\end{table*}

\begin{table}
\caption[]{Beam sizes and noise levels}
\label{tab:obs_freqs}
\begin{tabular}{l c c c c }
\hline\hline
Frequency  &  beam\tablefootmark{a} & \multicolumn{2}{c}{rms\tablefootmark{b}} \\
(GHz) &  ($''; \degr$) &  (mJy/Beam)& (K) \\
\hline
%199.9--201.8 & $4.36 \times   2.11$; 56 & 37 &  0.12  \\
%209.9--211.8 & $4.44 \times   2.33$; 54  & 35 &  0.093  \\
%201/211 cont. & $4.21 \times   1.86$; 56  & 4.4 & 0.016   \\
199.9--201.8 & $4.27 \times   1.84$; 57 & 38 &  0.15  \\
209.9--211.8 & $4.14 \times   1.95$; 55  & 37 &   0.13 \\
201/211 cont. & $4.14 \times   1.48$; 60  & 2.9 &  0.014  \\
\hline
344.1--345.9 & $0.45 \times  0.29$; 18 & 55 &  4.3  \\
354.1--355.9 &  $0.42 \times  0.29$; 22  & 55 & 4.4   \\
345/355 cont. &   $0.42 \times  0.28$; 20 & 12 &  1.0  \\
\hline
680.3--682.2 & $2.12\times 0.65$; 37 & 840 &  1.6  \\
690.3--692.2 &  $2.08\times 0.60$; 38 & 820 &  1.7  \\
681/691 cont. & $1.77 \times 0.50$; 42  & 227 &  0.67  \\
\hline
\end{tabular}
\tablefoot{
\tablefoottext{a}{Different weighting schemes were used for continuum and line images. Major and minor axis and position angle (from north to east) are given.}
\tablefoottext{b}{rms noise is given for a frequency width of 1~km~s$^{-1}$ in the line maps. It is evaluated from -10 to -5$''$ in both RA and Dec relative to the phase center.}
}
\end{table}

The high-mass star-forming region G10.47+0.03 was observed with the
Submillimeter Array\footnote{ The Submillimeter Array is a joint project
  between the Smithsonian Astrophysical Observatory and the Academia Sinica
  Institute of Astronomy and Astrophysics and is funded by the Smithsonian
  Institution and the Academia Sinica \citep{Ho04}.} (SMA) during five nights
in different array and receiver configurations, yielding six two-GHz-wide
bands centered at around 201, 211, 345, 355, 681, and 691
GHz. Table~\ref{tab:obs_dates} summarizes the observations. We note that
Jupiter was close ($\sim 10^\circ$) to the target source in 2008, so its moon
Callisto could be used as gain calibrator at 690~GHz, where it is otherwise
very difficult to find a suitable calibrator source. It was observed
repeatedly for 2~minutes after spending 10~minutes on the target source. For
the 345~GHz observations in 2007 and 2009, after each 15 minutes we switched
to 1733-130 (NRAO 530), which lies $10.9^\circ$ away from the target source. In addition, strong sources (Uranus, Ceres, 3c454.3, 3c273) were observed before and after these loops to allow bandpass and flux calibration.

%EX: 1924-292    uranus     3c454.3
%com: +  3c273 
%vex: +  1751+096    1911-201  

Calibration and editing of the data were done in the IDL MIR package\footnote{http://www.cfa.harvard.edu/∼cqi/mircook.html}, involving correction and application of the system temperature, phase-only bandpass calibration, continuum regeneration (spectral averaging), phase-and-amplitude bandpass calibration, flagging, gain calibration (phase and amplitude), and flux calibration. The data were converted to MIRIAD\footnote{http://bima.astro.umd.edu/miriad} \citep{Sault95}, where the edge channels of each chunk were flagged and the channels were rebinned to a width of 1 km~s$^{-1}$ at 201~GHz, 0.75 km~s$^{-1}$ at 345~GHz, and 1.42 km~s$^{-1}$ at 681/691~GHz. Data from different dates were merged for each frequency setup. Since the 2007 observations had the phase center at R.A. 18:08:38.28,  Dec. -19:51:50.0, while the later observations were centered on R.A. 18:08:38.232, Dec. -19:51:50.4, the 345/355~GHz data set had to be merged in AIPS. The channels less affected by spectral lines were identified and used for separating lines and continuum in MIRIAD. The continuum data of upper and lower sideband were merged and imaged using almost uniform weighting at 201/211~GHz and 345/355~GHz (robust$=-2$) and more natural weighting at 681/691~GHz (robust$=0.5$). A cutoff for cleaning of about 3 times the rms noise in the image was used. Based on these clean components, the continuum visibility data were self-calibrated, which substantially reduced the fluctuations outside the source, and the solutions were applied to the line data as well.  Imaging of the spectrally resolved data was done with robust$=0.5$ at 201/211~GHz and 345/355~GHz and robust$=2$ at 681/691~GHz. Table~\ref{tab:obs_freqs} gives the resulting beam sizes and noise levels in the maps. %In addition, line  to 5$''$ -> full spectra

Line identification was made with the myXCLASS program\footnote{https://www.astro.uni-koeln.de/projects/schilke/XCLASS}, which accesses the CDMS\footnote{http://www.cdms.de} and JPL\footnote{http://spec.jpl.nasa.gov} molecular data bases. All figures in this paper were made with the GILDAS software\footnote{http://www.iram.fr/IRAMFR/GILDAS}.

\section{Observational results}

\subsection{Continuum}

Figure~\ref{fig:cont} shows the obtained continuum maps. The total flux is 6 Jy at 201/211~GHz, 27 Jy at 345/355 GHz, and 95 Jy at 681/691 GHz, corresponding to a spectral index of 2.8 between the lower two and 1.8 between the upper two frequencies.  While the beam sizes are not sufficient to resolve the continuum emission at 201/211 and 681/691 GHz, the extension can be clearly seen at 345/355~GHz. At this frequency, the peak intensity is  158 K (1.84 Jy/Beam), and the $3\sigma$ contour extends over $3''$.

\begin{figure*}
  \centering
\includegraphics[bb=  85        343        510        744,angle=0,width=0.98\textwidth]{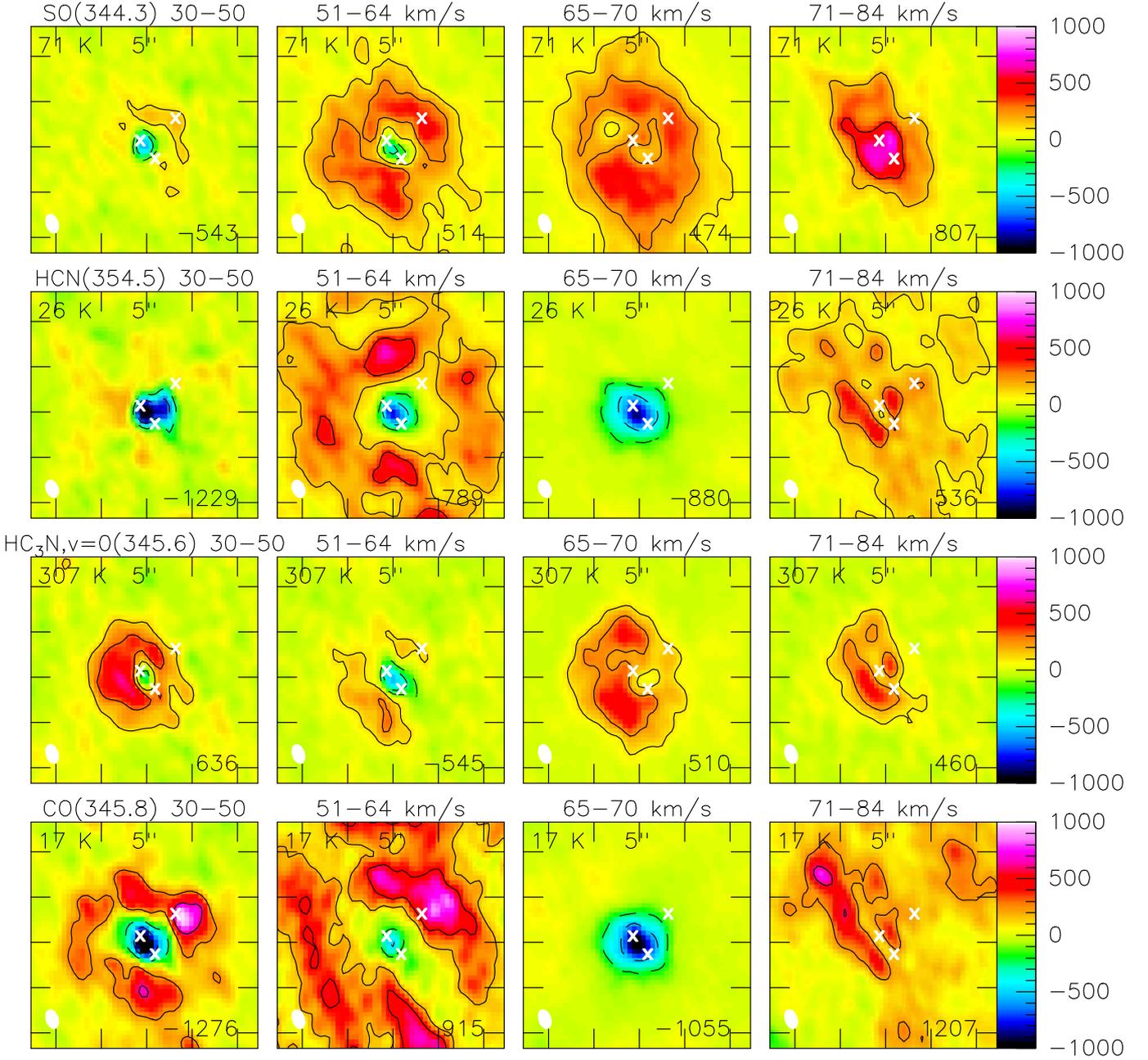}

 \caption{Line maps (selected to trace expansion motions) integrated
   over the velocity range indicated in each panel (which can be
   considered, from left to right, as high-velocity part of the
   outflow, low-velocity part, systemic velocity, and low-velocity
   part of the outflow at the far side).  The first velocity range of HC$_3$N is blended by other lines, in the second velocity range of CO the emission is so extended to cause imaging artifacts. The map size is 5$''$ (tick spaces are $1''$, centered on R.A. 18:08:38.236, Dec. -19:51:50.25). Beams are shown in the lower left, and the number in the lower right of each panel is the maximum flux in K~km~s$^{-1}$ (contours are $\pm$20 and 50\% of that value). The color scale is from -1000 to 1000  K~km~s$^{-1}$. The energy of the lower level is given in the upper left.}
  \label{fig:maps_of}
\end{figure*}

\begin{figure*}
  \centering
\includegraphics[bb=56        358        544        749,angle=0,width=0.98\textwidth]{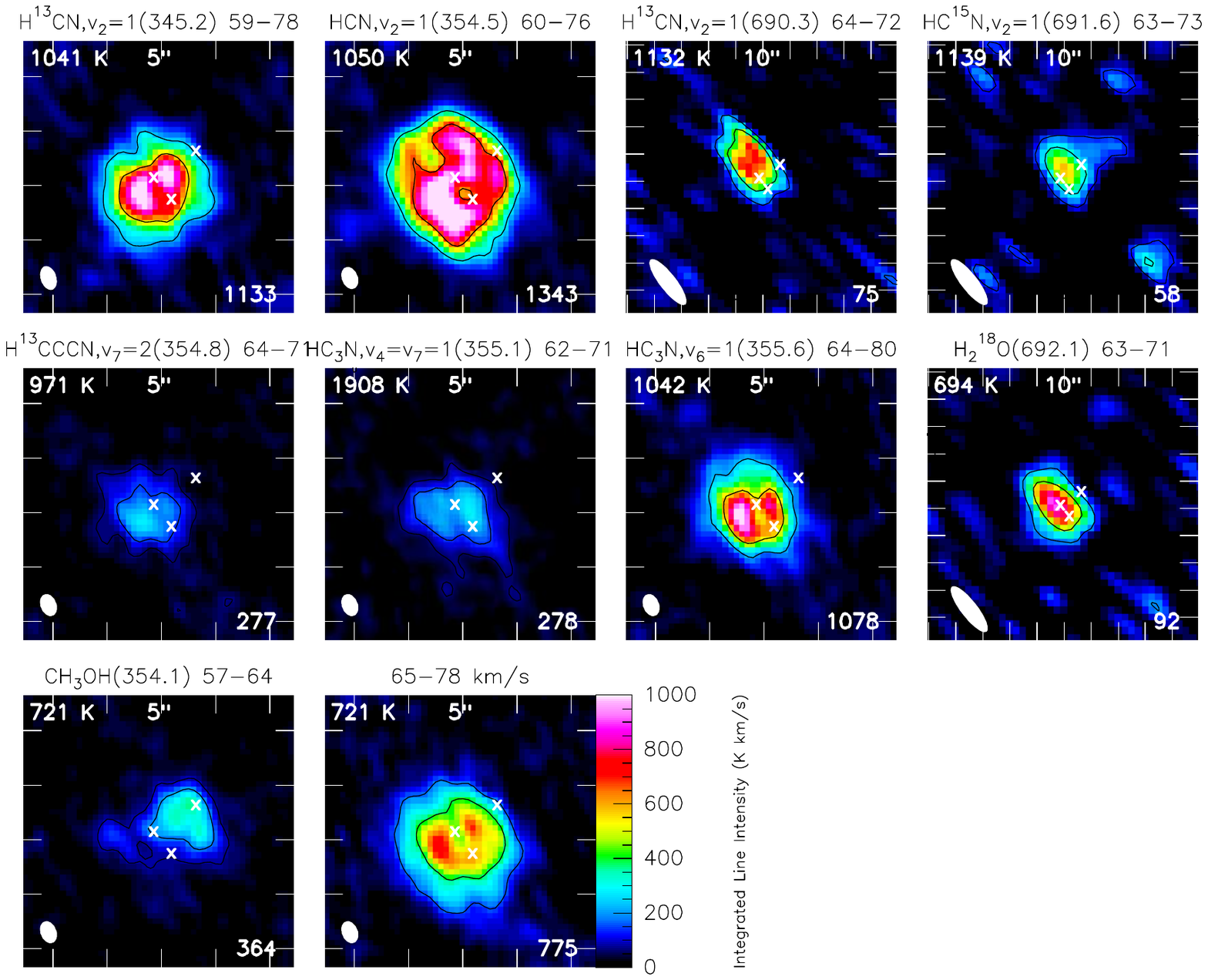}
 \caption{Selected high-excitation line maps.   Above each panel, the molecule, the frequency of the transition (in GHz), and the velocity range (in km~s$^{-1}$) are given. The color scale ranges from 0 to
   1000 K~km~s$^{-1}$ at 345/355 GHz,  where the map size is 5$''$,   and from 0 to 100
   K~km~s$^{-1}$ at 690 GHz, where the map size is 10$''$.  Beams are shown in the lower left, and the number in the lower right of each panel is the maximum flux in K~km~s$^{-1}$ (contours are $\pm$20 and 50\% of that value). The energy of the lower level is given in the upper left.}
  \label{fig:maps_he}
\end{figure*}

\subsection{Line identification}\label{sec:ident}

The data cubes were convolved to a common resolution of $5''$ and central
spectra were extracted. This resolution was chosen to decrease the
noise and the effects of a strong continuum, as well as to facilitate a
comparison between the different frequencies; it gives the total (source-integrated) flux for most lines. 

To identify the spectral features, a simple
homogeneous model was computed in Local Thermodynamic Equilibrium (LTE), using
the myXCLASS program, which takes the different optical depths of the transitions into account. Synthetic spectra are compared to the data. A source size of 1.5$''$ (half-maximum diameter of a Gaussian source), a temperature of 200 K and a
line width of 5 km~s$^{-1}$ were fixed, while the column density of each molecule  was varied to obtain a good fit to the data (see Table~\ref{tab:molecules}). The
continuum was neglected in the radiative transfer, only a foreground absorbing
column density of $5\times 10^{24}$ H$_2$ cm$^{-2}$  was taken into
account to approximate the higher absorption at the higher frequencies (corresponding to $\tau=0.3$ at 345 GHz and $\tau=1.3$ at 690 GHz). Although this is only foreground absorption, it serves to account for both absorption and continuum emission, as the latter weakens the lines as well.
Continuum levels of 6, 8, and 7 K were added to the spectra at
201/211, 345/355, and 681/691 GHz, respectively. Figures~\ref{fig:spec1},
\ref{fig:spec2}, and \ref{fig:spec3} show the spectra and the model, and Table~\ref{tab:molecules}
gives the derived column densities. We note that the model is not supposed to
give an optimum fit to the data, but is just for identification purposes.

\begin{table}
\caption[]{List of identified molecules}
\label{tab:molecules}
\begin{tabular}{l c c }
\hline\hline
Molecule & no. lines\tablefootmark{a} & column density\tablefootmark{b} \\
~ &  ~  &  $\left({\rm cm}^{-2}\right)$ \\
\hline
SO       &  5     &  4(17)  \\
SO$_2$   &   5    &  3(17)  \\
OCS   &   2    &    3(18) \\
\hline
CN   &   1    &   ...\tablefootmark{c} \\
HCN   &   8    &   1(18); 3(18)\tablefootmark{d} \\
HNC   &   1    &   3(17) \\
CH$_3$NH$_2$   & 2    &   4(17) \\
CH$_3$NC   &  4     &  7(15)  \\
NH$_2$CN   &  7     &   2(16) \\
NH$_2$CHO   &   15    &  2(17)  \\
HC$_3$N   &   76    & 5(16); 1(18)\tablefootmark{d} \\
C$_2$H$_3$CN   &  33     & 7(17)   \\
C$_2$H$_5$CN   &   55    &  9(17)  \\
\hline  
H$_2$O   &   1    &  1(20)  \\
CO   &    2   &   ...\tablefootmark{c} \\
H$_2$CO   &  6     &  3(18)  \\
CH$_3$OH   &  23     &  9(18)  \\
H$_2$C$_2$O   &  4     &  4(17)  \\
C$_2$H$_5$OH   &  23     &  6(17)  \\
CH$_3$OCH$_3$   &   14    &  1.5(18)  \\
%HCO$_2$H
CH$_3$OCHO   &  38     &   7(17) \\
CH$_3$CH$_3$CO   &   26    &  5(17)  \\
\hline
\end{tabular}
\tablefoot{The molecules are ordered by S-, N-, and O-bearing and by complexity.
\tablefoottext{a}{Number of identified lines (labeled in Figs.~\ref{fig:spec1}--\ref{fig:spec3})}
\tablefoottext{b}{Column density used in the myXCLASS model, with a source size of $1.5''$ and a temperature of 200 K. The parentheses are powers of 10. }
\tablefoottext{c}{The lines are in absorption, which is not modeled here.}
\tablefoottext{d}{For HCN and HC$_3$N, an additional component of 0.5$''$ and 500 K was used to more closely match the vibrational lines.}
}
\end{table}

%table with U-lines ?

\subsection{Line maps}\label{sec:linemaps}

From many maps of interesting lines (see appendix, Figs.~\ref{fig:maps1}--\ref{fig:maps5}), we show here only a selection, which are relevant for the expansion motion (Fig.~\ref{fig:maps_of}) and the high excitation (Fig.~\ref{fig:maps_he}). 
%Figures~\ref{fig:maps1}, \ref{fig:maps2}, and \ref{fig:maps3} show maps of a
%selection of interesting lines. 
They are ordered as in
Table~\ref{tab:molecules}.  The velocity was integrated over the ranges with
detectable signal and no significant deviations in the channel maps - i.e. the
whole flux for simple line shapes and several velocity ranges for more
complex line shapes.   For Fig.~\ref{fig:maps_of}, we chose common velocity ranges for the four lines, which are 30--50 km~s$^{-1}$, representing the high-velocity part of the (front-side) outflow, 51--64 km~s$^{-1}$, the low-velocity part, 65--70 km~s$^{-1}$, the systemic velocity, and 71--84 km~s$^{-1}$, the low-velocity part of the back-side outflow. Higher velocities are not detected, probably due to dust absorption. In case of contamination by a neighboring line, the
velocity range was chosen to avoid this line. Still, blending is likely for SO at 344.3 GHz by methanol (frequency corresponds to a 1.5 km~s$^{-1}$ lower velocity), H$^{13}$CN at 345.3 GHz by SO$_2$ (1 km~s$^{-1}$ higher velocity), H$^{15}$NC at 355.4 GHz by CH$_3$CH$_3$CO (3.3 km~s$^{-1}$ lower velocity), and CH$_3$NH$_2$ at 354.8 GHz by CH$_3$OCHO (3.8 km~s$^{-1}$ higher velocity).

%Integrated line maps: molecule, freq, range given in Fig.

%CH3OH at 64 km/s ?

\begin{figure*}
  \centering
  \includegraphics[angle=0,width=0.9\textwidth]{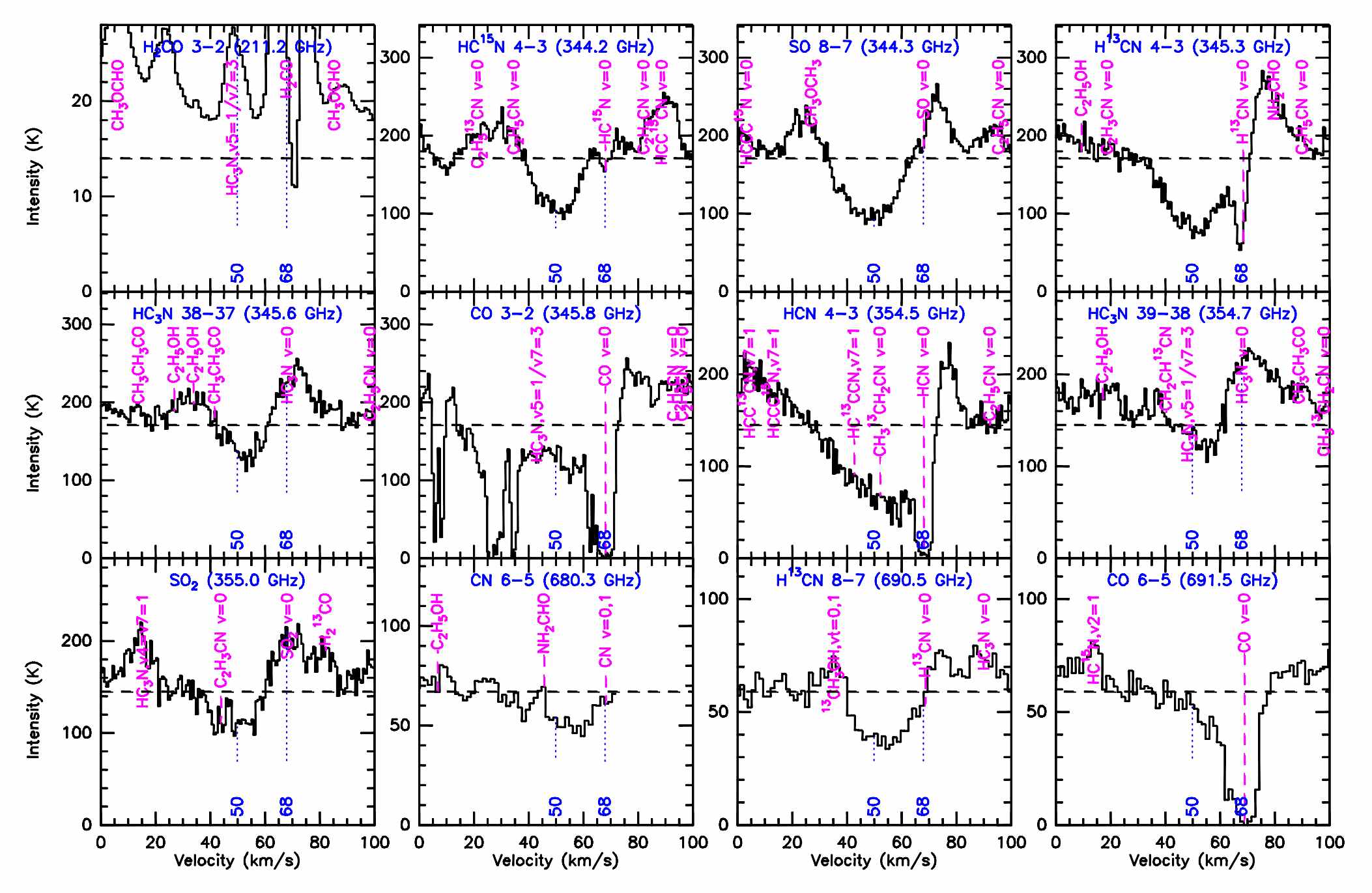} 
 \caption{Lines with absorption features toward the continuum
   peak. The systemic velocity of 68 km~s$^{-1}$ and a blue-shifted
   velocity of 50 km~s$^{-1}$ are marked. There are absorption components
   centered at both velocities, indicating expansion. The dashed horizontal line denotes the continuum level. }
  \label{fig:abslines}
\end{figure*}

\subsection{Spectra}

Figure~\ref{fig:abslines} shows central spectra of the 12 transitions with absorption features. Most of the absorption is blue-shifted relative to the systemic velocity of 68 km~s$^{-1}$. 

%Lines from highly excited HCN and HC$_3$N are shown in Figure~\ref{fig:highlines}. 

\begin{figure*}
  \centering
  \includegraphics[bb=51  263     567   758,angle=0,width=0.78\textwidth]{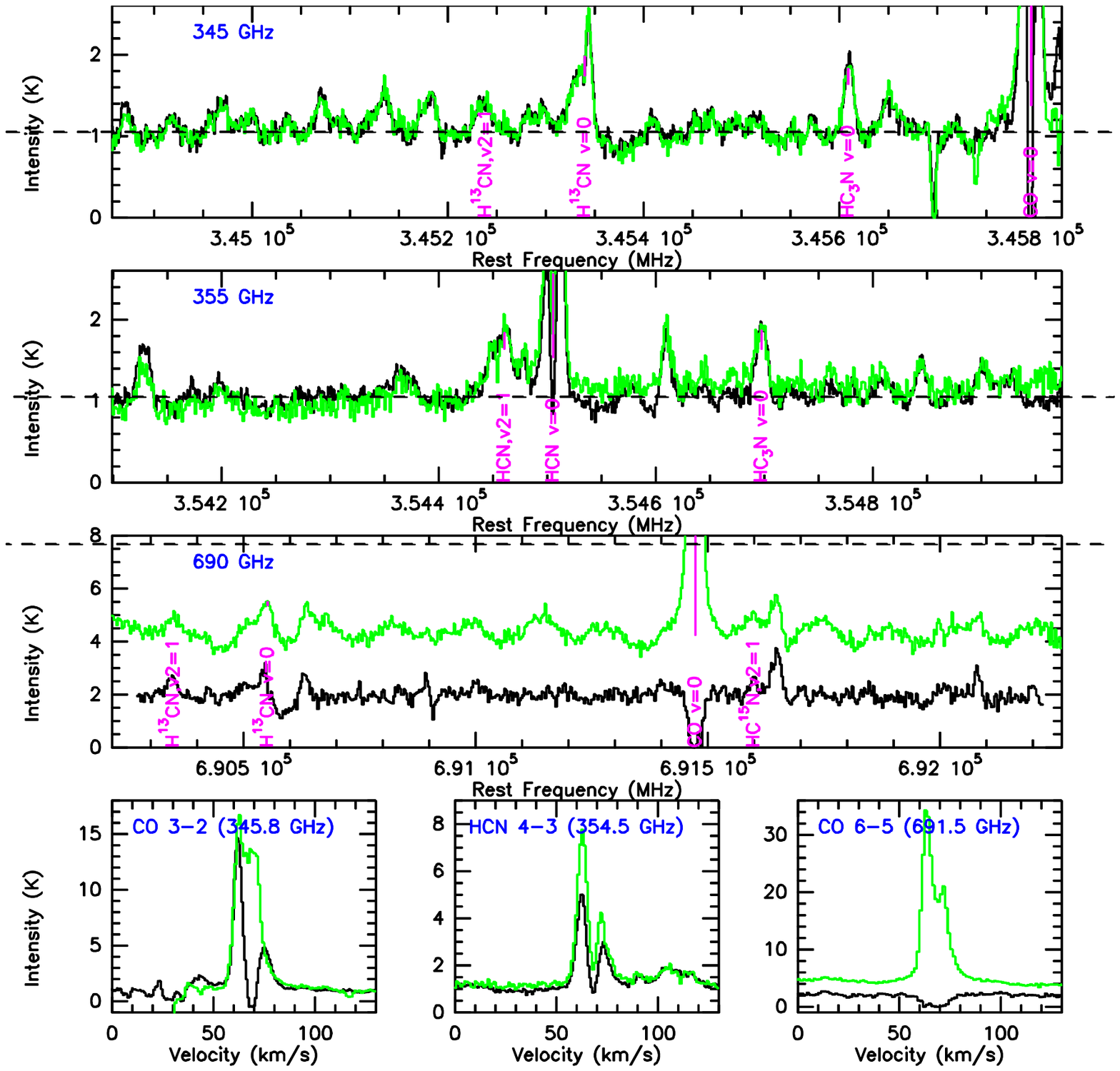} 
 \caption{Comparison of APEX data (green) to the SMA data convolved to
   the APEX beam (black). The dashed lines mark the continuum level as
   expected from LABOCA and SABOCA measurements. The HCN and CO lines
   and the 690~GHz continuum are affected by filtering of
   extended emission due to missing short spacings; the rest agrees
   very well. }
  \label{fig:apex}
\end{figure*}

\subsection{Comparison to APEX data}

To estimate calibration and filtering of extended emission, we compare the SMA
data to data from the APEX (Atacama Pathfinder Experiment) 12-m telescope
\citep{Guesten06,Schuller09,Rolffs11apex}. The 345 GHz flux from the LABOCA
bolometer array is 34 Jy/Beam, the 850 GHz flux from SABOCA is 450 Jy/Beam. At
both frequencies, the source is unresolved with the beams of 18.2 and 7.4$''$,
respectively. The spectral index is 2.86, so the 690 GHz flux would be 250
Jy. The SMA fluxes at 210 and 345 GHz fit very well to the APEX data, but at
690 GHz the continuum is much lower. This might be partly explained by
decorrelation due to fast phase fluctuations, and partly by missing short baselines filtering out extended emission.

%LABOCA (350 GHz) flux 34 Jy, consistent with point source     line cont. 30\%
%SABOCA (850 GHz) flux 450 Jy, consistent with point source         line cont. 15\%
%spectral index 2.9,  

%SMA 681/691 GHz flux should be 175 Jy, not 95 Jy!

Figure~\ref{fig:apex} shows APEX spectra overlaid with SMA spectra which were convolved to the APEX beam (18.2$''$ at 345 GHz, 17.7$''$ at 355 GHz, and 9.1$''$ at 690 GHz). The lines match  very well. The most notable exception is CO 6--5, which is purely in absorption with the SMA, but has strong emission as seen with APEX ($>$1000 Jy). Extended emission of CO 6--5 is filtered out due to missing short spacings, leaving an apparant absorption towards the continuum, which is more compact and hence not affected as much as CO by the filtering.

\section{Modeling}\label{sec:model}

In this section, we compare a few models to continuum and vibrationally
excited HCN in order to constrain the structure of this complex region, mainly
density distribution and velocity field. Testing different models of the source structure allows to exclude certain structures and to approach the real structure by adapting the models. The models are inevitably simplified,
and we used a trial-and-error technique with parameter variations and
comparison of model and data by eye to approach a good fit if possible for a
tested structure, but no global optimization of fit parameters was performed.

The three-dimensional radiative-transfer code RADMC-3D\footnote{http://www.ita.uni-heidelberg.de/\textasciitilde dullemond/software/radmc-3d}, developed by C.~Dullemond, was employed to compute the dust temperature from stellar heating and the continuum and line emission of a model. The basic setup is the same as described in \citet{Rolffs11VLA}, but here we also test clumpiness and velocity structure in the models. The line transfer assumes LTE, which is a good approximation for vibrationally excited HCN, but prevents the modeling of the ground-state lines.

\begin{table*}
\caption[]{Summary of the models  described in Sect.~\ref{sec:models}.}
\label{tab:models}
\begin{tabular}{l c c  c c}
\hline\hline
Model &  density  & heating &  velocity field  & displayed in Figs.\\
\hline
A  & power-law  &  1 star  &  infall   & \ref{fig:cont_models}, \ref{fig:spec_AB} (red) \\
B  & Gaussian   &  3 stars &  none & \ref{fig:cont_models}, \ref{fig:spec_AB} (green) \\
C  & Plummer    &  3 stars &  none  & \ref{fig:cont_models} (blue), \ref{fig:spec_CD} (red) \\
D  & clumpy &  3 stars &  dispersion & \ref{fig:cont_models} (cyan), \ref{fig:spec_CD} (green) \\
E  & clumpy, outflow cone &    3 stars  & dispersion, outflow    &  \ref{fig:cont_models} (magenta), \ref{fig:model_cont}, \ref{fig:model_vla}-\ref{fig:model_h13cn8-7}  (red) \\ 
\hline
\end{tabular}
\end{table*}

\subsection{The Models}\label{sec:models}

While no perfect fit was found, we selected five of the tested models for presentation and comparison to the data. Models A and B have been previously used to fit APEX and VLA data of this source, respectively \citep{Rolffs11apex,Rolffs11VLA}. Model C has a radial density profile that best matches the observed continuum (if heated by the stars in the H{\sc ii} regions). Models D and E have density fluctuations on small scales, which are expected from self-gravity and which improve the line fitting. Model E is an attempt to include the outflow (which is seen by the blue-shifted absorption features) and is presented in more detail.

The models are described explicitly in the following. Table~\ref{tab:models} summarizes the main properties of the models.

\paragraph{Model A} This is the model that was presented in \citet{Rolffs11apex} to fit the APEX data (continuum and many lines from HCN, HCO$^+$, and CO, including vibrationally excited HCN). The density follows a radial power law, $n=7\times 10^8\times \left( \frac{r}{485 {\rm AU}} \right)^{-1.75}$ H$_2$ cm$^{-3}$ for radii larger than 485 AU. Inside there is an H{\sc ii} region with an electron density of $1.5\times 10^6$ cm$^{-3}$ to reproduce the free-free radiation from B1, and a star of $5.6\times 10^5$ L$_\odot$. The dust opacity is from \citet{Ossenkopf94} without grain mantles, but with coagulation at a density of $10^5$ cm$^{-3}$. The HCN abundance is $3\times 10^{-5}$ at temperatures above 100 K and $5\times 10^{-8}$ below 100 K. The line width is 5 km~s$^{-1}$ FWHM, and the gas is radially infalling with 1 km~s$^{-1}$ to the center at B1. The total mass (in a cube of 3 pc diameter) is $2.4\times 10^4$ M$_\odot$, of which 70 M$_\odot$ are at temperatures above 300 K.

\paragraph{Model B} This is the model that was presented in \citet{Rolffs11VLA} to fit the VLA data of vibrationally excited HCN. The density follows a Gaussian, centered at B1, with  $7\times 10^7$ H$_2$ cm$^{-3}$ at the half-maximum radius of 7000 AU. As in all following models, heating sources are the stars in the H{\sc ii} regions B1 with $10^5$ L$_\odot$, B2 with $8.3\times 10^4$ L$_\odot$, and A with $4.6\times 10^4$ L$_\odot$, which are placed in the plane of the sky.
. The dust opacity is from \citet{Ossenkopf94} without grain mantles or coagulation, as in all following models. The HCN abundance is $10^{-5}$. The intrinsic line width (FWHM) is 8.3 km~s$^{-1}$, and there is no macroscopic velocity field. The total mass is  $3.5\times 10^3$ M$_\odot$, of which 350 M$_\odot$ are at temperatures above 300 K.

\paragraph{Model C} The density in this model follows a Plummer profile, $n=1.4\times 10^8 \times 
\left(1 + \left(\frac{r}{11500 {\rm AU}}\right)^2\right)^{-2.5}$ H$_2$ cm$^{-3}$ (half-maximum radius 6500 AU). It is centered 2000 AU south and 1000 AU west of B1. The Plummer model is very similar to a Gaussian inside the half-maximum radius, but has higher density outside (still falling off steeply as $r^{-5}$). The HCN abundance is $10^{-5}$ at temperatures above 300 K and $10^{-6}$ below, as in the following models. The intrinsic line width (FWHM) is 8.3 km~s$^{-1}$, and there is no macroscopic velocity field. The total mass is  $6.8\times 10^3$ M$_\odot$, of which 350 M$_\odot$ are at temperatures above 300 K.

\paragraph{Model D}  This model consists of 100 clumps, which have a Plummer
half-maximum radius of 1000 AU and central densities ranging from  $4\times
10^8$  to  $2\times 10^9$  H$_2$ cm$^{-3}$. The distribution of clump masses follows the same slope as the stellar Initial Mass Function. They are randomly placed in the model according to a Plummer
distribution with half-maximum radius 4000 AU. The density in each model cell
is the maximum of all contributions from the clumps. So with a lot of overlap,
this structure consists rather of density fluctuations than of separate cores.
The intrinsic line width (FWHM) is 5 km~s$^{-1}$. Each clump has a random
line-of-sight velocity with Gaussian half-maximum value of $\pm 5$
km~s$^{-1}$. The line-of-sight velocity of each cell is an average of all clumps, weighted with their density contribution. The total mass is  $9.4\times 10^3$ M$_\odot$, of which 470 M$_\odot$ are at temperatures above 300 K.

\begin{figure}[h!]
  \centering
  \includegraphics[width=0.49\textwidth]{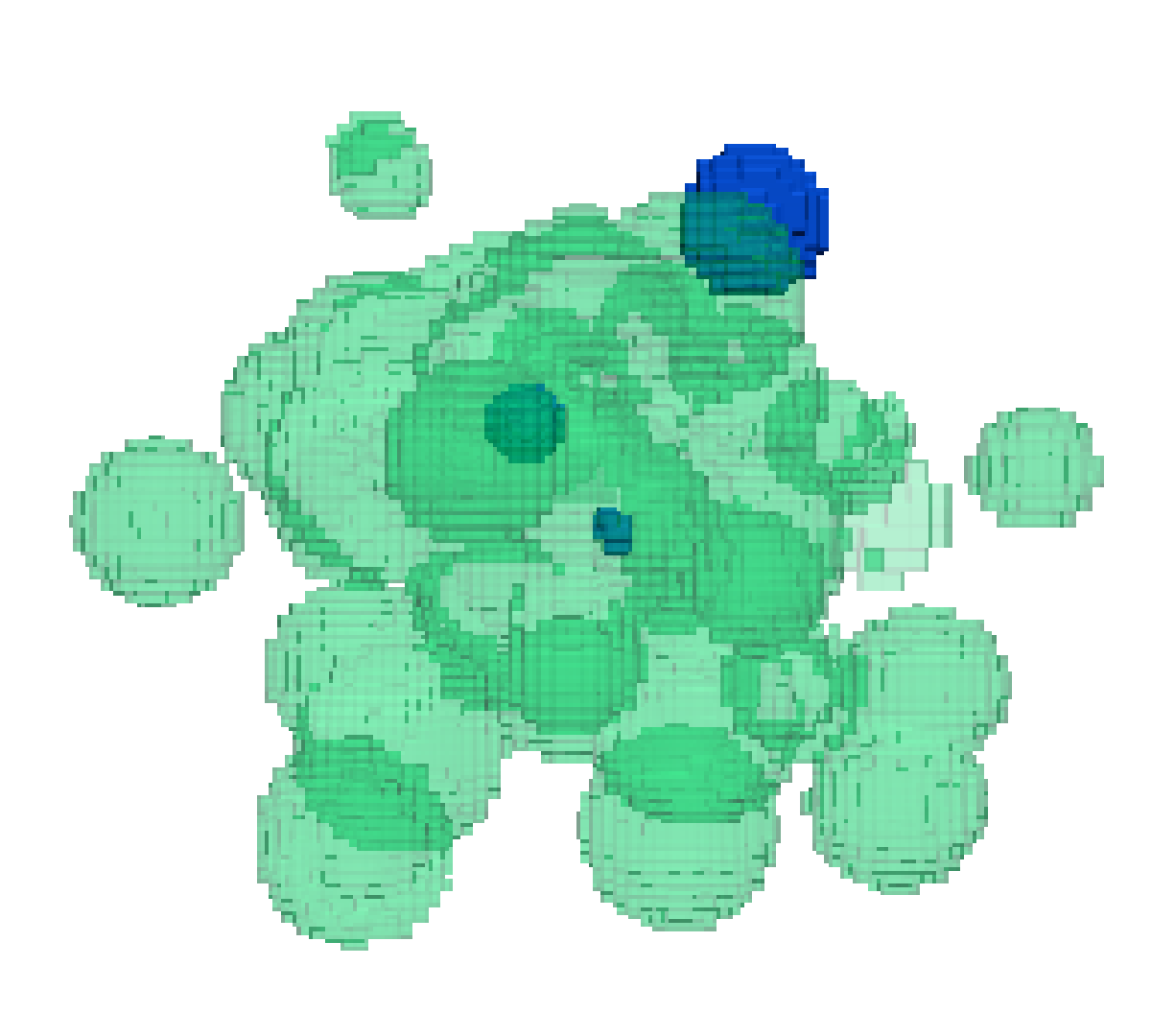}  %contours.pdf} 
 \caption{Clumpy structure of model E displayed as isocontours of $10^7$ H$_2$  cm$^{-3}$ (green, transparent). The H{\sc ii} regions, whose stars heat up the gas, are shown in blue. The size of the contoured region is around 0.2 pc.}
  \label{fig:clumpy}
\end{figure}

\paragraph{Model E}  This model has the same structure as model D, but a
half-maximum radius of the distribution of the clumps of only 3000 AU. In
addition, a bipolar outflow is present, with an opening angle of 100$^\circ$
and a length of $2\times 10^4$ AU. Its center lies 1000 AU south and 500 AU
west of B1. The axis is tilted from the line of sight by 20$^\circ$, and the
foreground part directed towards B2. Inside the outflow cone, the density is
reduced to 20\% of the original value (thus preserving the clumpy structure) and the velocity is 10 km~s$^{-1}$ outwards. This is only a
toy model of the outflow. The total mass is  $4.6\times 10^3$ M$_\odot$, of which 400 M$_\odot$ are at temperatures above 300 K. The clumpy structure is visualized in Fig.~\ref{fig:clumpy}.

\subsection{Continuum}

The continuum radiation that the models emit is compared to the data. Models D and E have randomly placed core centers; therefore their radiation (continuum and lines) have a random component, which cannot be exactly reproduced in different runs with the same parameters.

\begin{figure}[h!]
  \centering
  \includegraphics[angle=0,width=0.49\textwidth]{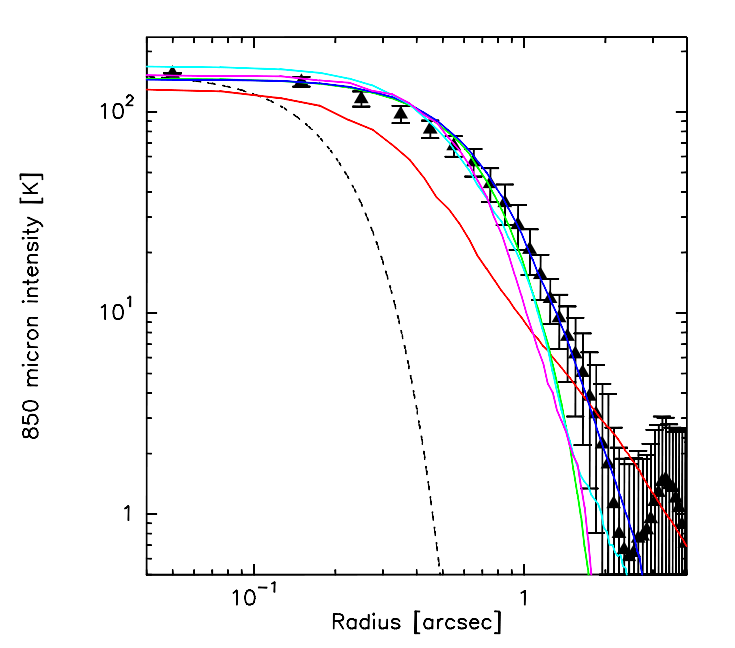} 
 \caption{Models A (red), B (green), C (blue), D (cyan), and E
   (magenta) compared to the SMA continuum profile, which is displayed
 as data points with a $0.1''$ radial interval. The errorbars denote the rms
 deviation from a circular shape, and the dashed curve represents the beam.}
  \label{fig:cont_models}
\end{figure}

\begin{figure*}
  \centering
  \includegraphics[width=0.89\textwidth]{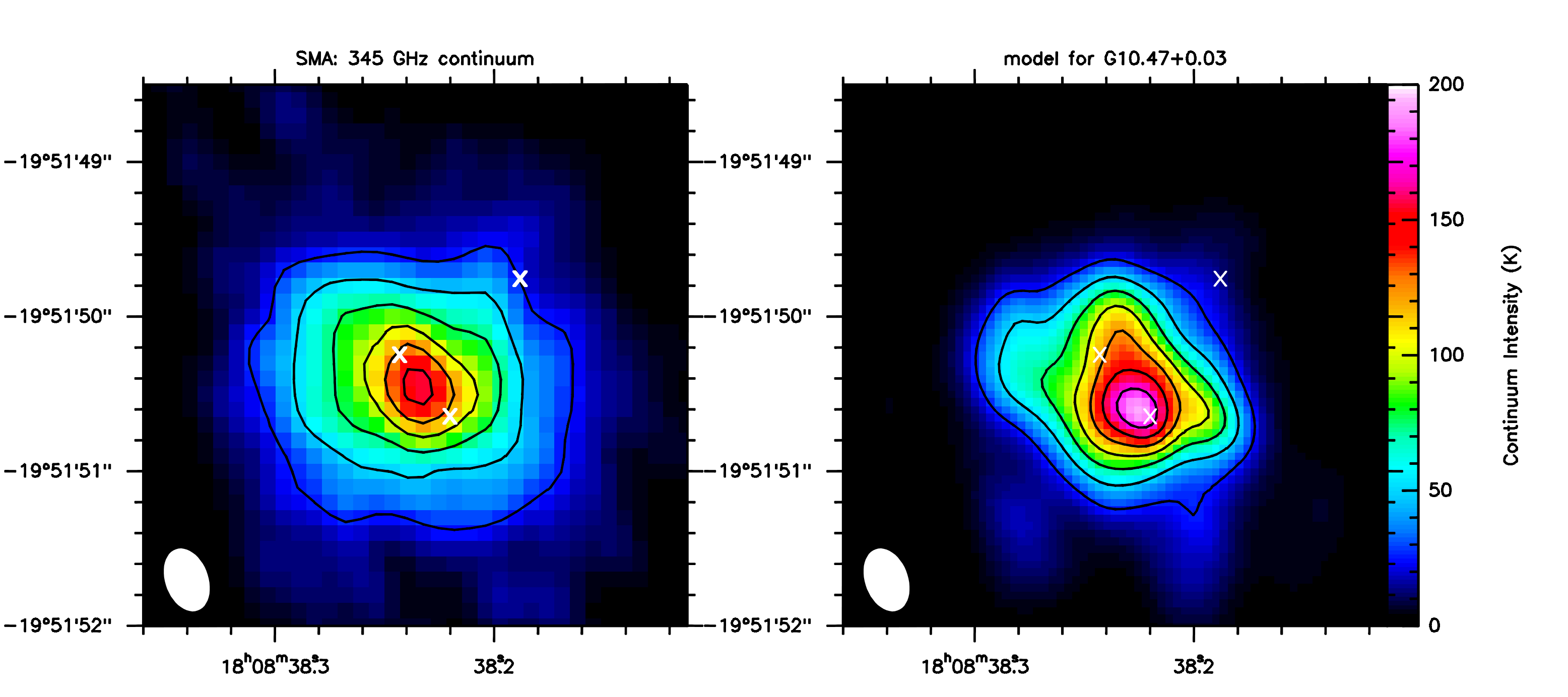} 
 \caption{Model E (right) compared to the SMA continuum map at 345~GHz
   (left). }
  \label{fig:model_cont}
\end{figure*}

The radial profile of the 345/355~GHz continuum map is extracted with central
position R.A. 18:08:38.237, Dec. $-19$:51:50.421, which is the center of a
two-dimensional Gaussian fit ($1.4''\times 1.17''$, elongated along the B1-B2
axis). The model maps were Fourier-transformed, folded with the
uv-coverage of the observations, and imaged in the same way as the data. The
radial profile of the model was extracted from the same central
coordinates. Figure~\ref{fig:cont_models} shows a comparison of the radial
profiles, Fig.~\ref{fig:model_cont} a comparison of the continuum map of model
E to the data.  Model A (power-law) fits worse to the radial profile, model C (Plummer) fits best.

%b2n: radial continuum profile extracted with central position R.A. 17:47:19.889, Dec. -28:22:18.317, which is the center of a two-dimensional Gaussian fit ($1.94''\times 1.52''$, elongated north-south)

The flux from APEX/LABOCA (345 GHz) is 34 Jy/Beam ($18.2''$ beam size) and
from APEX/SABOCA (850 GHz) 450 Jy/Beam (7.4$''$ beam size). The models emit 31
(A), 20 (B), 26 (C), 20 (D), and 18 (E) Jy/Beam for LABOCA and 266 (A), 231
(B), 311 (C), 146 (D), and 162 (E) Jy/Beam for SABOCA. In models B-E an
extended component similar to A could be added without affecting the results of the interferometer modeling.

%LABOCA flux: data 34 Jy/Beam, APEX model 31, VLA model 20  , Plummer 26 , clumps 20 , clumps+outflow 18.
%SABOCA flux: data 450 Jy/Beam, APEX model 266, VLA model 231  , Plummer 311 ,clumps 146 , clumps+outflow 162  

\begin{figure}
  \centering
  \includegraphics[bb=53        124        445        722,angle=0,width=0.49\textwidth]{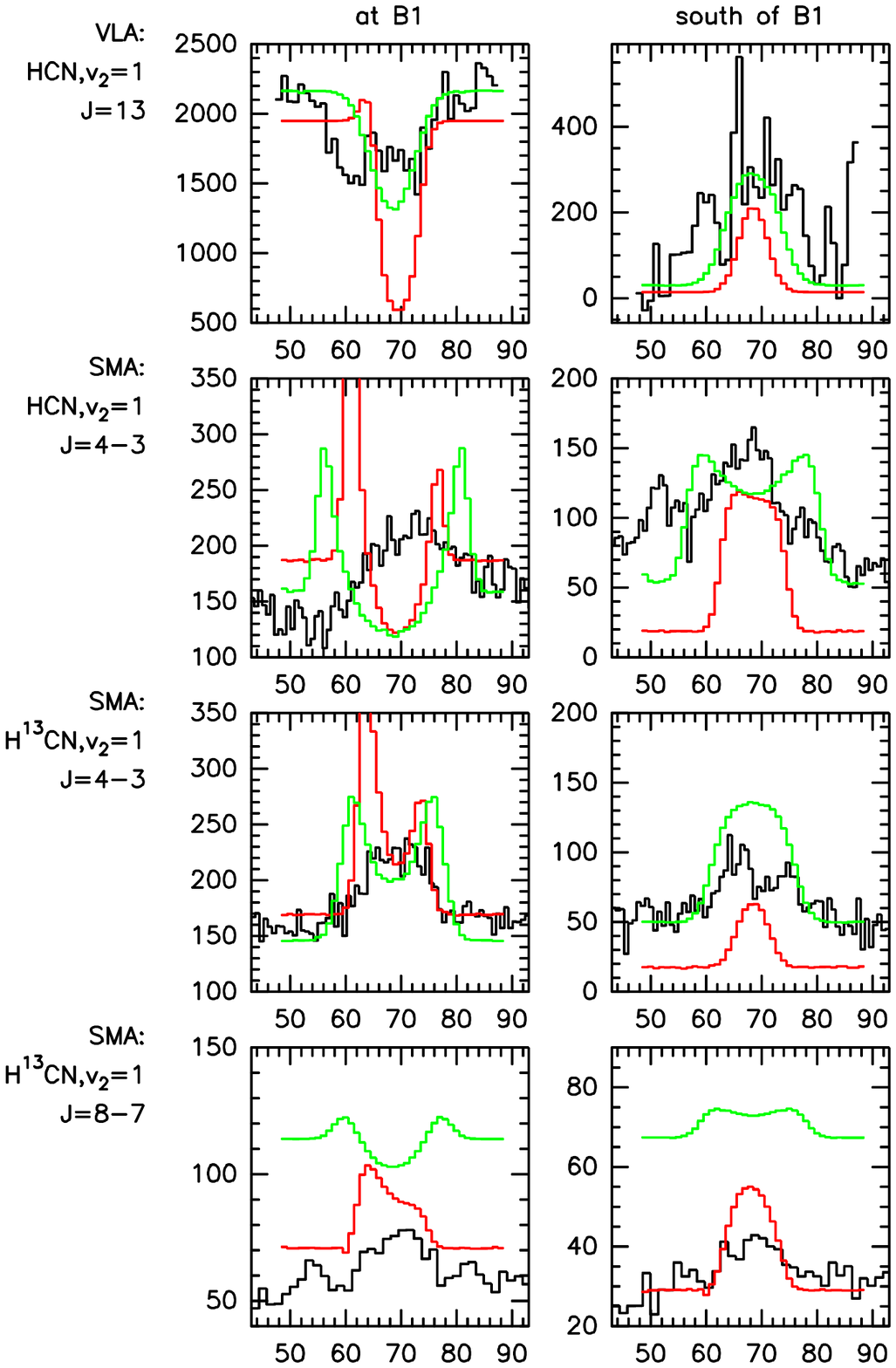} 
 \caption{Models A (red) and B (green) compared to lines from vibrationally excited HCN at the position of B1 and $0.3''$ (VLA) or $0.8''$ (SMA) south of B1.}
  \label{fig:spec_AB}
\end{figure}

\begin{figure}
  \centering
  \includegraphics[bb=53        124        445        722,angle=0,width=0.49\textwidth]{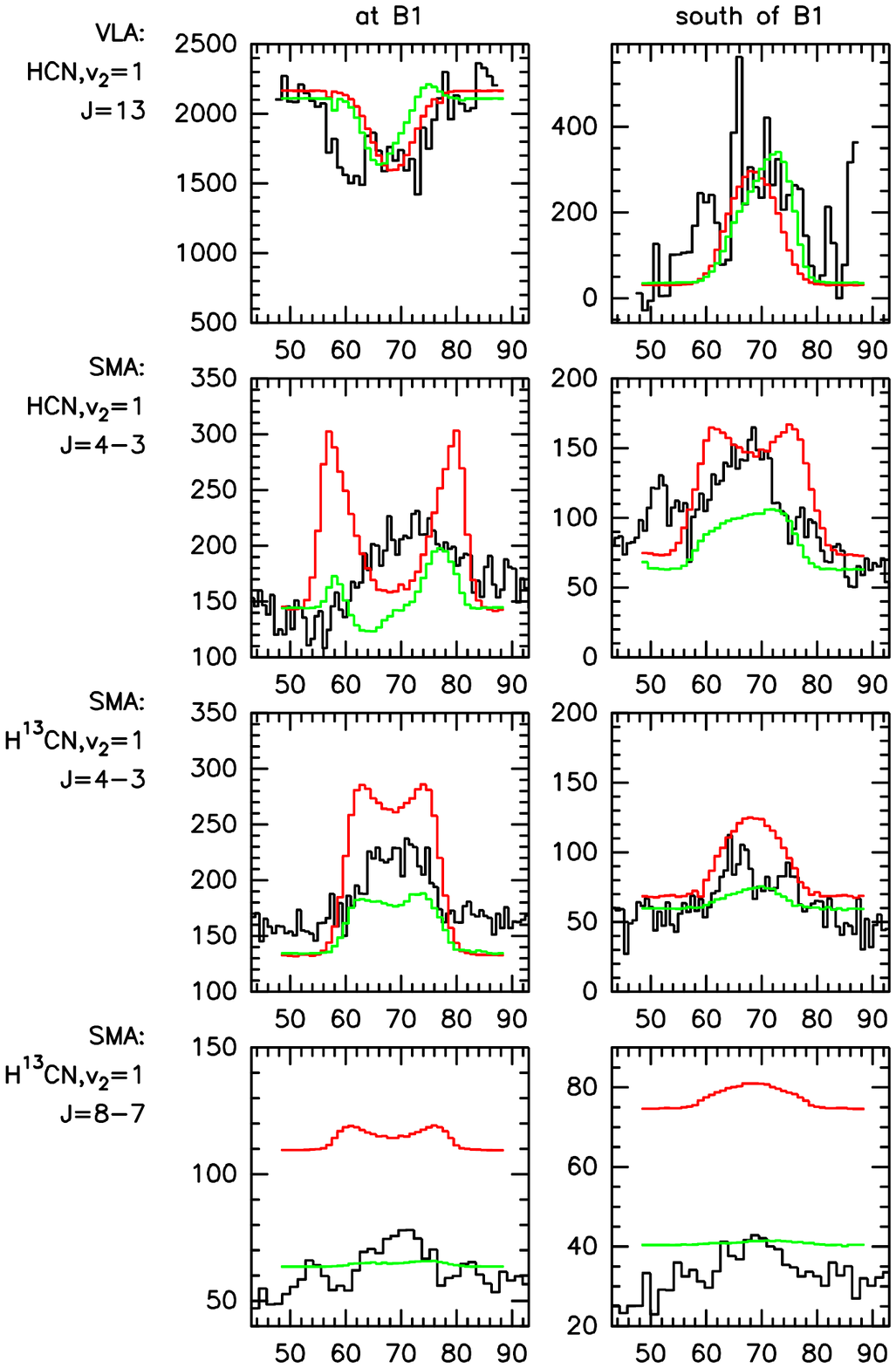} 
 \caption{Models C (red) and D (green) compared to lines from vibrationally excited HCN at the position of B1 and $0.3''$ (VLA) or $0.8''$ (SMA) south of B1.}
  \label{fig:spec_CD}
\end{figure}

\begin{figure*}
  \centering
  \includegraphics[width=0.89\textwidth]{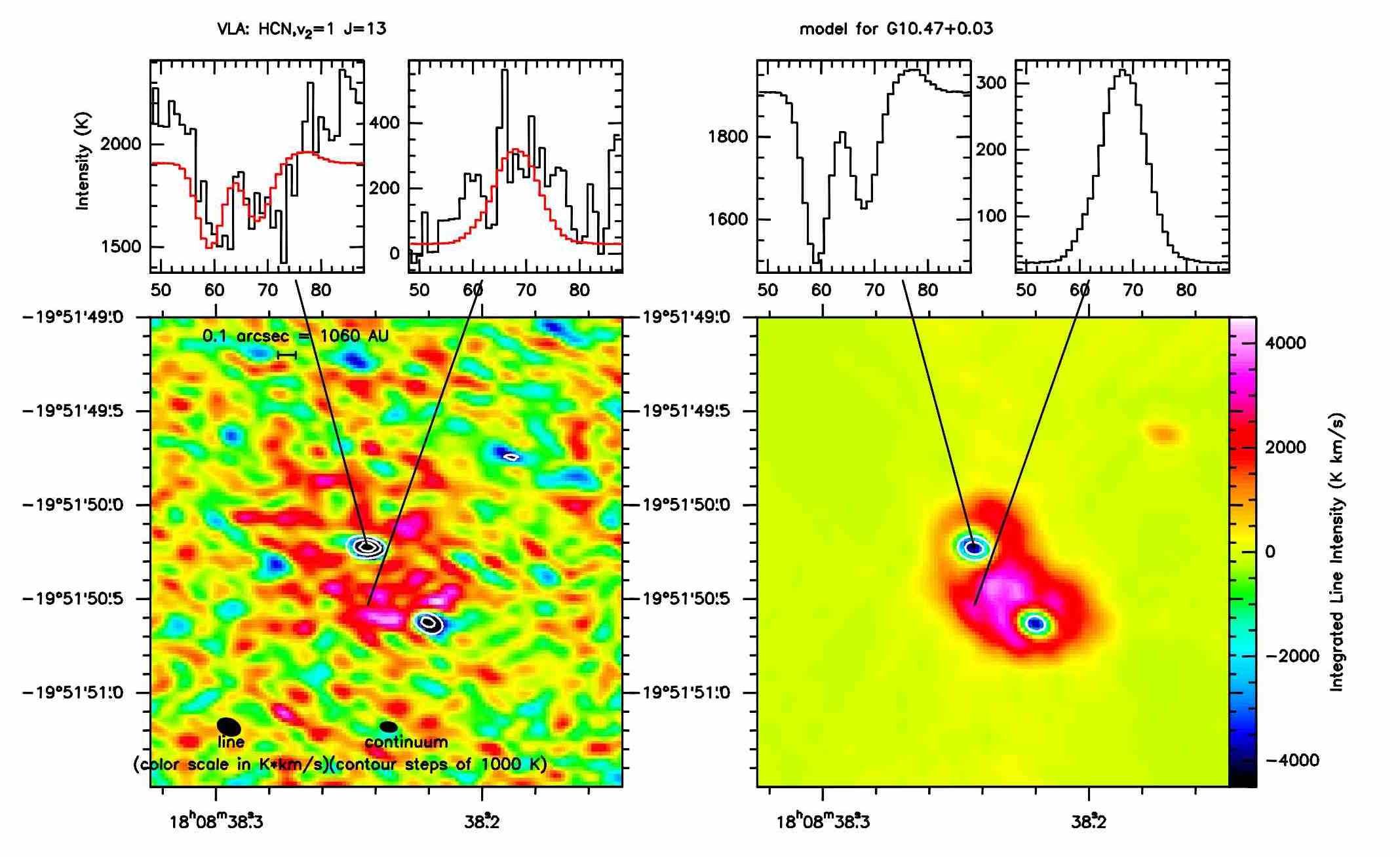} 
 \caption{Model E (right) compared to the $J$=13 direct $\ell$-type line of
   vibrationally excited HCN at 40.8 GHz (VLA), shown as integrated
   line map and spectra at two locations (left, model spectra are overlaid in red). The white contours denote the 7~mm continuum in steps of 1000 K. }
  \label{fig:model_vla}
\end{figure*}

\begin{figure*}
  \centering
  \includegraphics[width=0.89\textwidth]{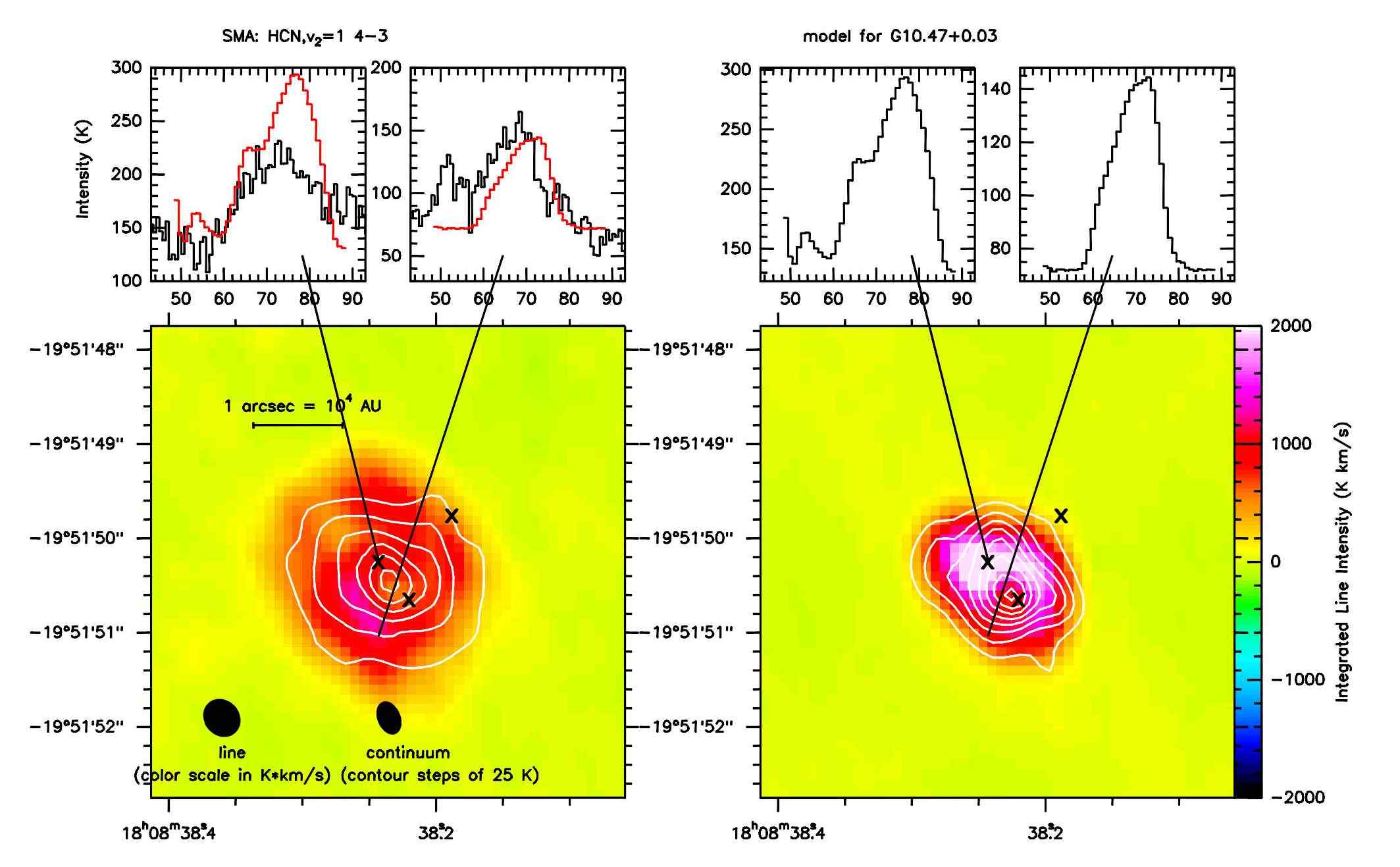} 
 \caption{Model E (right) compared to the $J$=4--3 line of vibrationally excited HCN at 354.5 GHz (SMA), shown as integrated
   line map and spectra at two locations (left, model spectra are overlaid in red). The white contours denote the 355~GHz continuum in steps of 25 K. }
  \label{fig:model_hcn4-3}
\end{figure*}

\begin{figure*}
  \centering
  \includegraphics[width=0.89\textwidth]{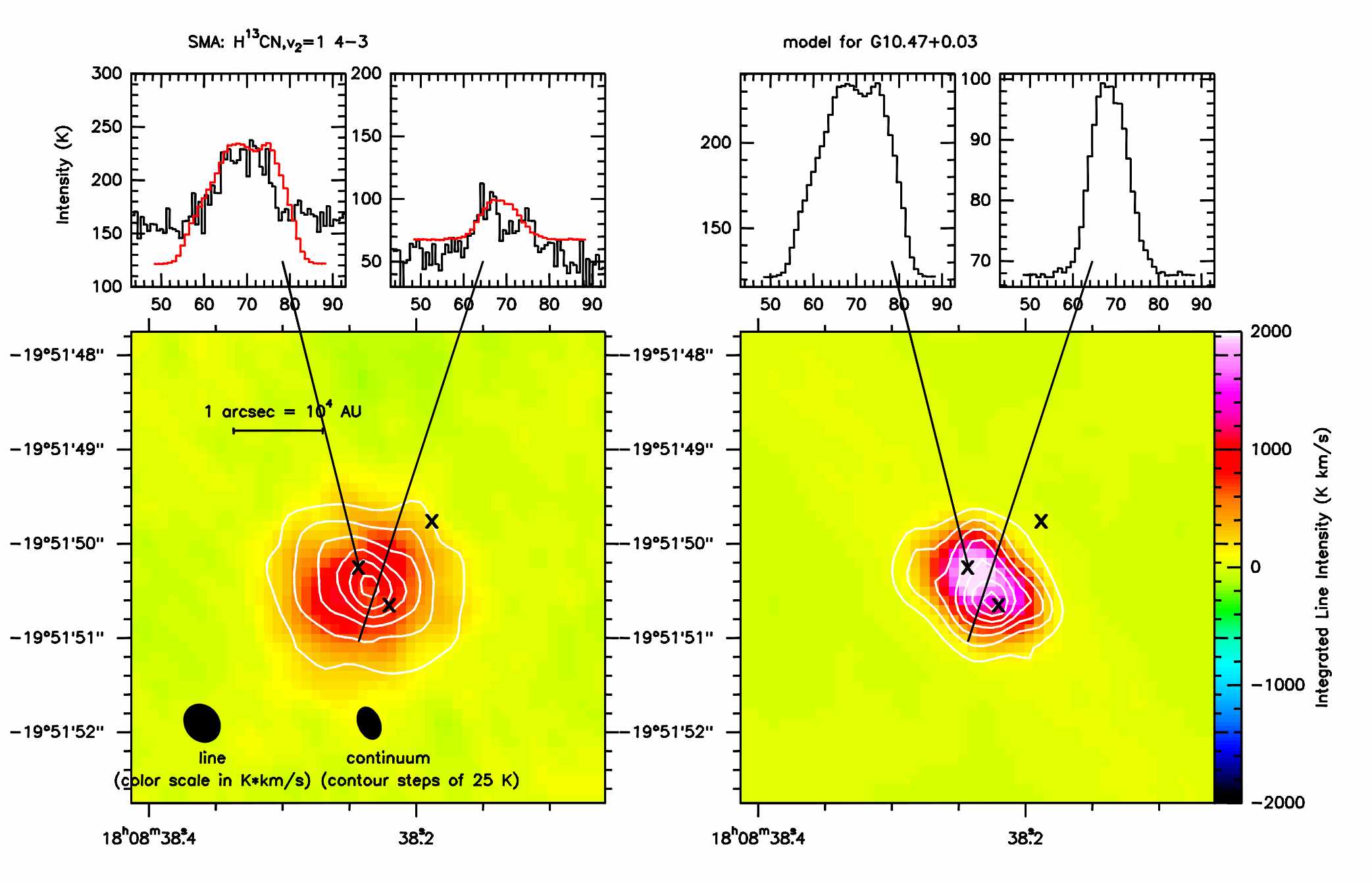} 
 \caption{Model E (right) compared to the $J$=4--3 line of vibrationally excited H$^{13}$CN at 345.2 GHz (SMA), shown as integrated
   line map and spectra at two locations (left, model spectra are overlaid in red). The white contours denote the 345~GHz continuum in steps of 25 K.  }
  \label{fig:model_h13cn4-3}
\end{figure*}

\begin{figure*}
  \centering
  \includegraphics[width=0.89\textwidth]{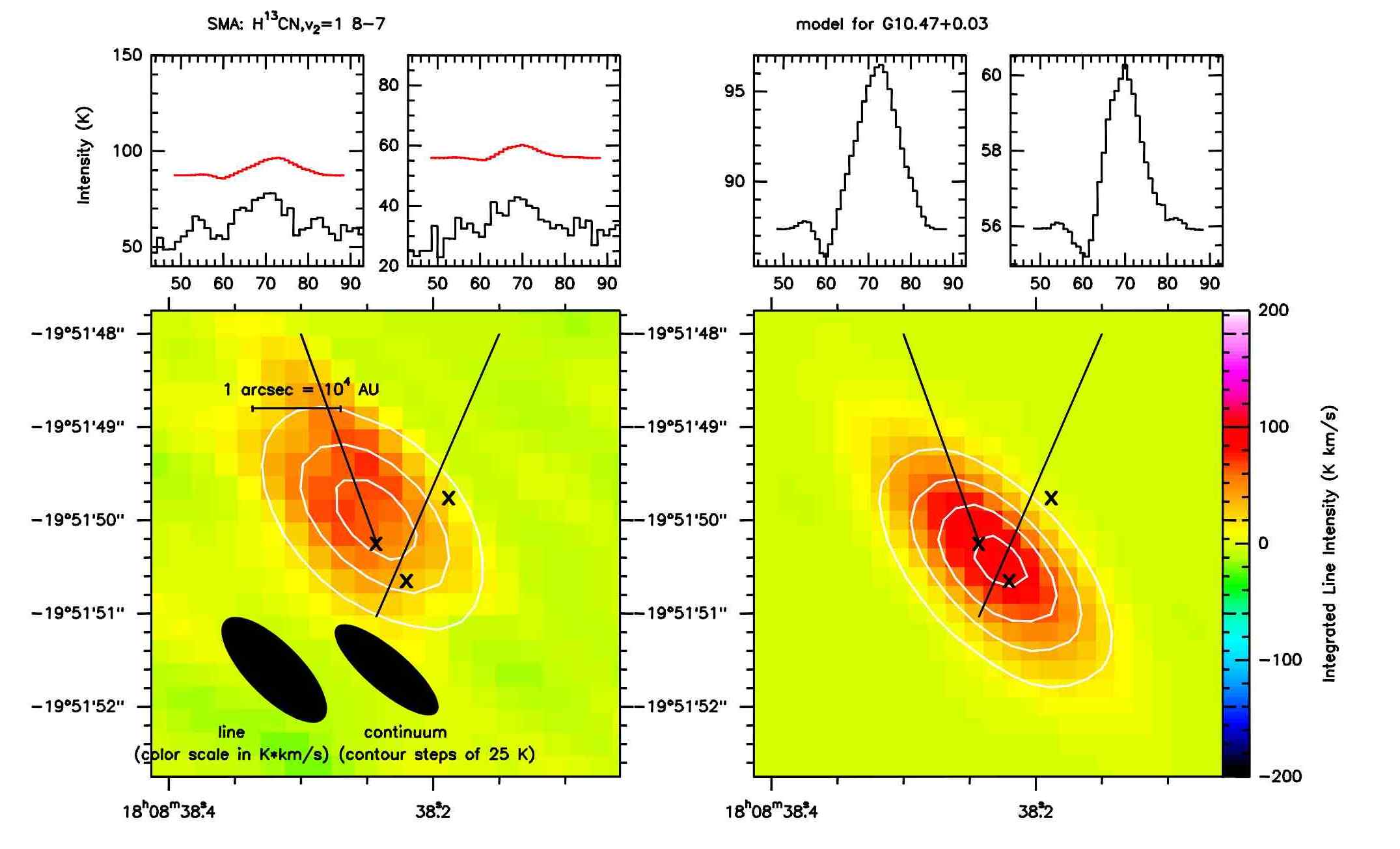} 
 \caption{Model E (right) compared to the $J$=8--7 line of vibrationally excited H$^{13}$CN at 690.4 GHz (SMA), shown as integrated
   line map and spectra at two locations (left, model spectra are
   overlaid in red). The white contours denote the 690~GHz continuum in steps of 25 K.  }
  \label{fig:model_h13cn8-7}
\end{figure*}

\subsection{Vibrationally excited HCN}

The models are compared to lines from vibrationally excited HCN, including the $J$=13 direct $\ell$-type transition at 40.7669 GHz observed with the VLA \citep{Rolffs11VLA}. Figures~\ref{fig:spec_AB} and \ref{fig:spec_CD} show this line and three rotational transitions of vibrationally excited HCN observed with the SMA, overlaid with models A--D.  To reproduce the direct $\ell$-type line, the rotational lines must be very optically thick, leading to self-absorption in the models. This is inevitable for optically thick lines from centrally heated spheres; the data however do not show self-absorption.  The self-absorption decreases from model A to E. Model E is shown in more detail in comparison to the $J$=13 direct $\ell$-type line of vibrationally excited HCN (Fig.~\ref{fig:model_vla}), the 4--3 line of vibrationally excited HCN (Fig.~\ref{fig:model_hcn4-3}) and H$^{13}$CN (Fig.~\ref{fig:model_h13cn4-3}), and the 8--7 transition of vibrationally excited H$^{13}$CN (Fig.~\ref{fig:model_h13cn8-7}).  The outflow is imprinted in the line shapes. The levels of all these transitions lie between 1050 and 1400 K above ground.

%\subsection{HC$_3$N}

%maybe more molecules: myCloud?

\section{Discussion}

\subsection{Chemistry}

In addition to about 350 identified lines, there are around 90 unidentified
(U-) lines. 75 of these 90 U-lines are in the 201/211 GHz range, which covers
a similar velocity range as the higher-frequency bands combined (almost 6000
km~s$^{-1}$). This is probably due to the lower noise at these frequencies. The line density is mainly determined by the width of single
lines (5--10 km~s$^{-1}$), i.e. a forest of overlapping lines covers most of
the bands. The identified molecules (3 S-bearing, 10 N-bearing, and 9 other,
O-bearing, molecules) represent only the strongest lines, while more complex
molecules have many weak lines which probably add up to the continuum and are
undetectable due to line confusion. No spatial separation of different molecules could be seen. The differences in the maps can probably be explained by excitation and optical depth.

In the detailed models  (Sect.~\ref{sec:model}), the abundance of HCN in the hot gas is very high,
on the order of $10^{-5}$  relative to H$_2$. This is needed to reproduce the direct $\ell$-type line and the vibrationally excited H$^{13}$CN. Around 5\% of the nitrogen and 1.4\% of the carbon is in HCN \citep[for solar N/H of $1.1\times 10^{-4}$ and C/H of $3.6\times 10^{-4}$,][]{Anders89}, and HCN/CO is on the order of 0.1 in the hot central region. We note that such a high HCN abundance needs an explanation by future chemical models, incorporating high-temperature reaction networks.
The abundance of HC$_3$N seems to follow a similar
increase at high temperatures, as the strong emission from vibrational states suggests (Figs.~\ref{fig:spec1}--\ref{fig:spec3}). 
Notable line detections are HC$^{15}$N,$v_2$=1
at 691.6 GHz (lower level is 1139 K above ground), HC$_3$N,$v_4$=$v_7$=1 at
355.1 GHz (1908 K), and the $^{13}$C substituted  HC$_3$N,$v_7$=2 (970 K). The
  H$_2^{18}$O line at 692.1 GHz (694 K) implies a high abundance of
  water in the dense, warm gas. These abundance enhancements could be connected, e.g. through a reduced abundance of OH, which at high temperatures reacts to form water instead of destroying N-bearing molecules \citep{Rodgers01}.

\subsection{Density Distribution}

As the high-resolution continuum map at 345/355 GHz shows, a power-law radial
density is not consistent with the data. A Gaussian fits well in the inner
part, but falls off too steeply in the outer parts. What fits best to the data
is a Plummer model (Fig.~\ref{fig:cont_models}, blue). This profile is very similar to a Gaussian inside the
half-maximum radius, but is denser outside, where it falls off as $r^{-5}$. It
is also used to describe the stellar density of star clusters. Although that
might reveal a connection of this forming star cluster to later stages, it
could also have completely different physical reasons. The stellar density
distribution is determined by stellar dynamics, while the gas density
distribution is determined by the interplay of different sources of pressure
(gravitational versus turbulent, rotational, magnetic, radiative, and thermal
pressure). The observed central flattening of the density can be explained by
centrally increased pressure, which is expected from feedback by the newly
formed massive stars.  This pressure stalls the infall and piles up the mass where the infall stops, resulting in a less centrally peaked mass distribution.

The SMA continuum at 690~GHz is rather weak compared to the 345~GHz
flux and the high-frequency single-dish flux. As interferometer
observations at these frequencies are challenging and only very few
have been conducted yet \citep[e.g.][]{Beuther06}, this could be due
to decorrelation caused by rapid phase noise. On the other hand, the
line strengths are similar to the APEX data (Fig.~\ref{fig:apex}),  supporting the data quality and the calibration, so
if the low continuum is real, it can be due to optical depth effects
in very dense, small condensations, while extended emission is
filtered out. The continuum at  690~GHz must be more extended than at
345~GHz because the larger dust opacity makes it stronger only in the
optically thin regions, while it can be weaker in the optically thick
case  - instead of the whole hot interior, one sees only up to the colder foreground. In addition,  the image fidelity at  690~GHz is not as good as at 345~GHz, in particular the shortest baselines  are  twice as large. As the flux rises with shorter baselines, this means much more filtering out of extended emission at 690~GHz.

There is no fragmentation observed in neither the continuum nor the lines. The reason may be that we do not resolve structures smaller than the beam size of about 3000 AU. We note that the continuum source extends over about 10 beams ($3''$), so no large-scale fragmentation is present. Models of the line emission reveal that a spherical density distribution leads to strong self-absorption features, which are not observed. Clumpiness better reproduces the line shapes, but without sufficient spatial resolution we cannot derive the properties of fragmentation.

This missing fragmentation is similar to the contiuum data recently obtained by \citet{Qin11} for SgrB2-N, which follows a very similar radial profile and also displays a similar spectral line content. Maybe the two sources are at the same evolutionary stage, where the first hypercompact H{\sc ii} regions have developed and heated the gas, but large-scale fragmentation has not set in.

The rotational transition of vibrationally excited HCN (4--3 at 354.46 GHz)
 has a very high optical depth. The Einstein A coefficient is 50,000 times larger
 than for the direct $\ell$-type line ($J$=13 at 40.7669 GHz, observed with
 the VLA), which has an optical depth of $\lesssim 1$. Also the comparison to the H$^{13}$CN line (Fig.~\ref{fig:maps_he}) shows its high optical depth. In models which are homogeneous on small scales (smooth density gradient, such as models A--C), this optically thick line is  self-absorbed due to the temperature gradients caused by internal heating.   Such large temperature gradients are inevitable as the stars must be deeply embedded, shown by the high efficiency of heating and the compactness of the H{\sc ii} regions.
The observed single-peaked profile can only be explained by
an inhomogeneous density and velocity field  (such as models D--E), where the self-absorption is smeared out. Such small-scale clumpiness might persist in later stages of evolution, when the gas is ionized \citep{Ignace04}. 
 %, since the temperature gradient otherwise leads to a strong double-peaked profile.

%\citet{Pfalzner09}

The observed continuum depends on density, temperature, and dust opacity. The density can vary by many orders of magnitude, and therefore dominates the resulting continuum. It is clear that the temperature is high in the inner region and falls off outwards; this is included in the models as the stars in the hypercompact H{\sc ii} regions heat the dust. It is entirely possible, however, that more heating sources are present, although they are not needed for heating the central region.  The extension of high-excitation lines is generally a bit smaller in the models than in the data, especially highly excited HC$_3$N is somewhat extended to the east (Fig.~\ref{fig:maps_he}).

\subsection{Velocity Field}

While most lines do not show any velocity structure in their maps, there seems
to be a component of highly excited CH$_3$OH at lower velocity close to the
UCH{\sc ii} region A (Fig.~\ref{fig:maps_he}). Alternatively, this could arise from
blending by a different line.

On large scales, the asymmetries of self-absorbed lines clearly indicate
infall motions \citep[][and Fig.~\ref{fig:apex}]{Rolffs11apex}, and also the absorption of the H$_2$CO line is a bit red-shifted (Fig.~\ref{fig:abslines}). On smaller
scales, we see expansion motions, as very nicely traced by the absorption
features (Fig.~\ref{fig:abslines}). The blue-shifted absorption is stronger in the western
part of the continuum. In addition, the SO line at 344.3 GHz, the
HCN line at 354.5 GHz, and the CO line at 345.8 GHz have red-shifted
emission in the north-eastern part. A red-shifted clump  $5''$ north of the hot
core, which is only seen in HCN and CO, seems to be too far away to be related
to the outflow (Fig.~\ref{fig:maps_of}).

\citet{Olmi96} detected a north-south velocity gradient in $^{13}$CO, and
\citet{Hofner96} see blue-shifted water masers $0.2''$ north and $0.4''$ south
of B1, and a red-shifted water maser (at 70 km~s$^{-1}$) $1.2''$ north of B1.
Absorption against the hypercompact H{\sc ii} regions B1 and B2 is found to be
blue-shifted in NH$_3$(4,4) \citep[][53 km~s$^{-1}$ toward B1 and 61
  km~s$^{-1}$ toward B2]{Cesaroni10}, and also partly in vibrationally excited HCN \citep[][and Fig.~\ref{fig:model_vla}]{Rolffs11VLA}.

\begin{figure}
  \centering
  \includegraphics[angle=0,width=0.49\textwidth]{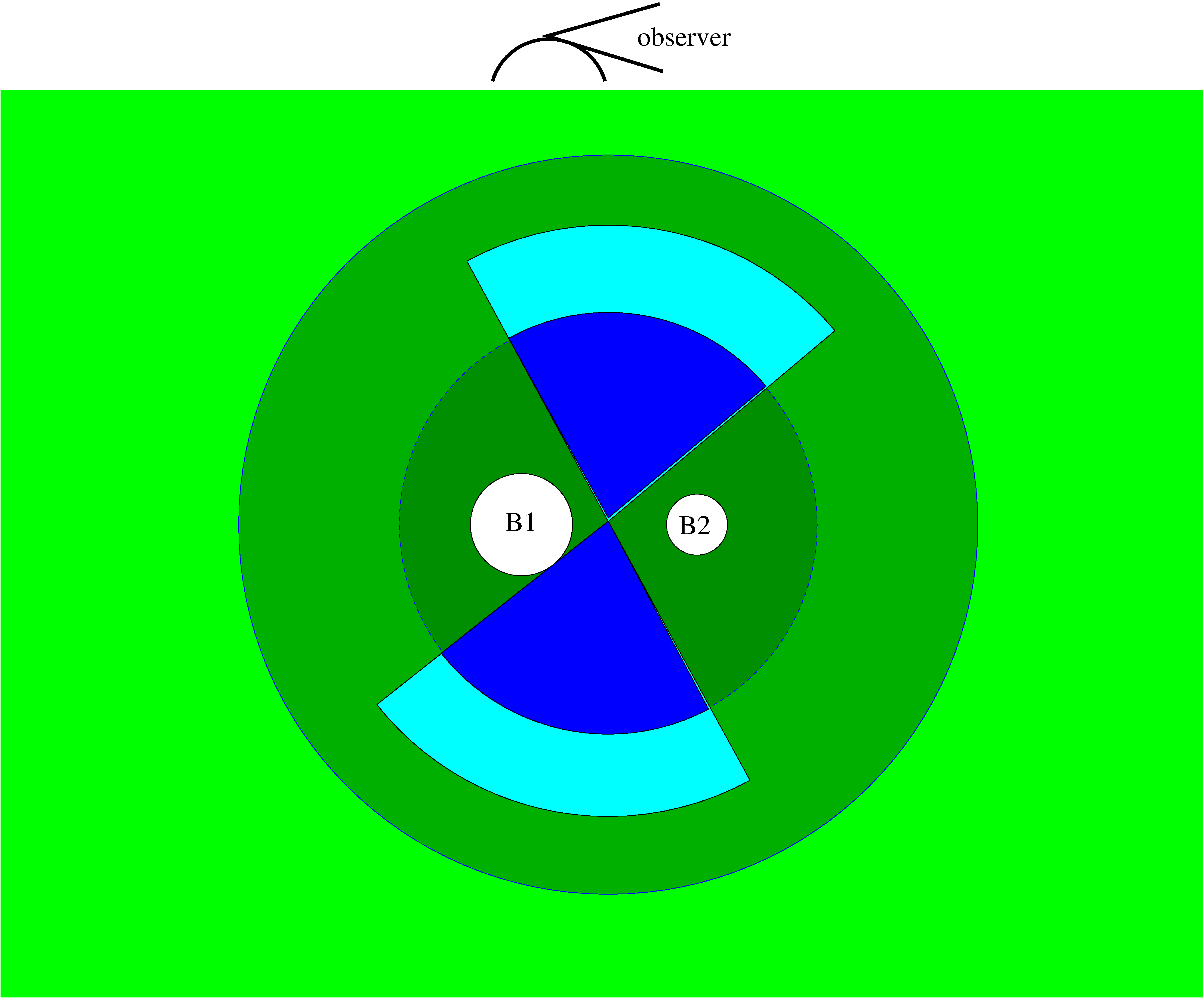} 
 \caption{Qualitative scenario for the outflow in G10.47+0.03. The observer is above the plot. Infalling molecular gas is shown in green, expanding molecular gas in blue, and ionized gas in white.  The circles denote half-maximum and Plummer radius in Model C (the gas must be clumpy, though) and the boundaries of the H{\sc ii} regions, which have density gradients  \citep{Cesaroni10}. If the H{\sc ii} regions are not spherical, the outflow could also originate from the ionizing stars.   }
  \label{fig:sketch}
\end{figure}

One can conclude that there is likely a bipolar outflow, whose foreground part
is tilted to the south-west, and whose background part is
tilted to the north-east. A sketch of the outflow
scenario is presented in Fig.~\ref{fig:sketch}. This scenario explains
qualitatively all observed velocity features. The different NH$_3$(4,4) velocities, for instance, can be understood as an effect of the projection of the outflow to the line-of-sight. An embedded source, which has not yet developed a detectable H{\sc ii} region, is driving the outflow in this scenario. A nearly spherical infall with lower velocity, but higher rate than the expansion surrounds the outflow and produces the blue asymmetric line profiles seen on larger scales. This accretion probably continues well into the central region along certain paths, driven by gravitational attraction and the momentum of the large-scale accretion flow.

An alternative scenario might arise if the hypercompact H{\sc ii} regions B1 and B2 are not spherical, but in the form of bipolar expanding bubbles from a photoevaporating disk \citep{Hollenbach94}. In this case, the stars in the HCH{\sc ii} regions would drive two outflows, which are both roughly aligned with the line-of-sight.

In addition, it is possibile that the source is  a large pseudo-disk 
accreting from a massive envelope, rotating roughly in the plane of the 
sky, driving an outflow perpendicular to it, and having formed B1 and B2 as  
binaries (and maybe more stars). This would correspond to the transition 
between phase II and III of \citet{Zapata10}.

 A quantitative radiative transfer model, which reproduces all observed lines, is possible, but beyond the scope of this paper. That the outflow is important for reproducing the line shapes has
been demonstrated in our modeling, as only model E could reproduce the shapes (Sect.~\ref{sec:model}).

\section{Conclusions}

With the SMA, we have obtained high-resolution, spectrally resolved maps of the massive hot molecular core G10.47+0.03 at different frequencies, covering a bandwidth of 12~GHz in total and including observations at 690~GHz. Our main results are

\begin{itemize}

 \item Hundreds of molecular lines reveal a rich chemistry, with molecules such as HCN and HC$_3$N especially abundant at high temperatures.  This high abundance demands explanations from chemical models. Vibrationally excited HCN, whose levels lie more than 1000 K above ground, is very optically thick, and even vibrationally excited HC$^{15}$N shows up at 690 GHz.

 \item Blue-shifted absorption in a dozen lines indicates an outflow oriented roughly along the line-of-sight. It is embedded as there is also absorption at the systemic velocity of cold foreground gas in CO and HCN. 

 \item The averaged radial profile of the submm continuum displays a central flattening and rapid falloff, and is best fitted by a Plummer profile of the density. The mass of the source is on the order of several thousand M$_\odot$, of which a few hundred are at high ($>$300 K) temperatures.

 \item No fragmentation is observed over the $\sim$10 beam sizes (30,000 AU) of the continuum emission. High-excitation lines which are very optically thick do not show self-absorption. The line modeling suggests that this must be due to density fluctuations in combination with the velocity field. If the rather low continuum at 690 GHz is real, it points to small, very dense condensations.

\end{itemize}

From these findings, a picture emerges of a very young forming star cluster, characterized by beginning of feedback from massive stars, while infall is ongoing. This feedback includes heating of the dust and gas, ionization that is still confined to small regions, and increased pressure in the inner part that leads to expansion motions and central flattening of the density.

\begin{acknowledgements}
We are grateful to Mark Gurwell for his help with the reduction of the 690 GHz
data, and to Kees Dullemond for kindly providing us with the RADMC-3D
code. R. Rolffs acknowledges support from the International Max Planck
Research School (IMPRS) for Astronomy and Astrophysics.
\end{acknowledgements}

\bibliographystyle{aa}
\bibliography{ref}

\Online

\begin{appendix}

\section{Full Spectral Range}

\begin{figure*}
  \centering
  \includegraphics[angle=0,width=0.99\textwidth]{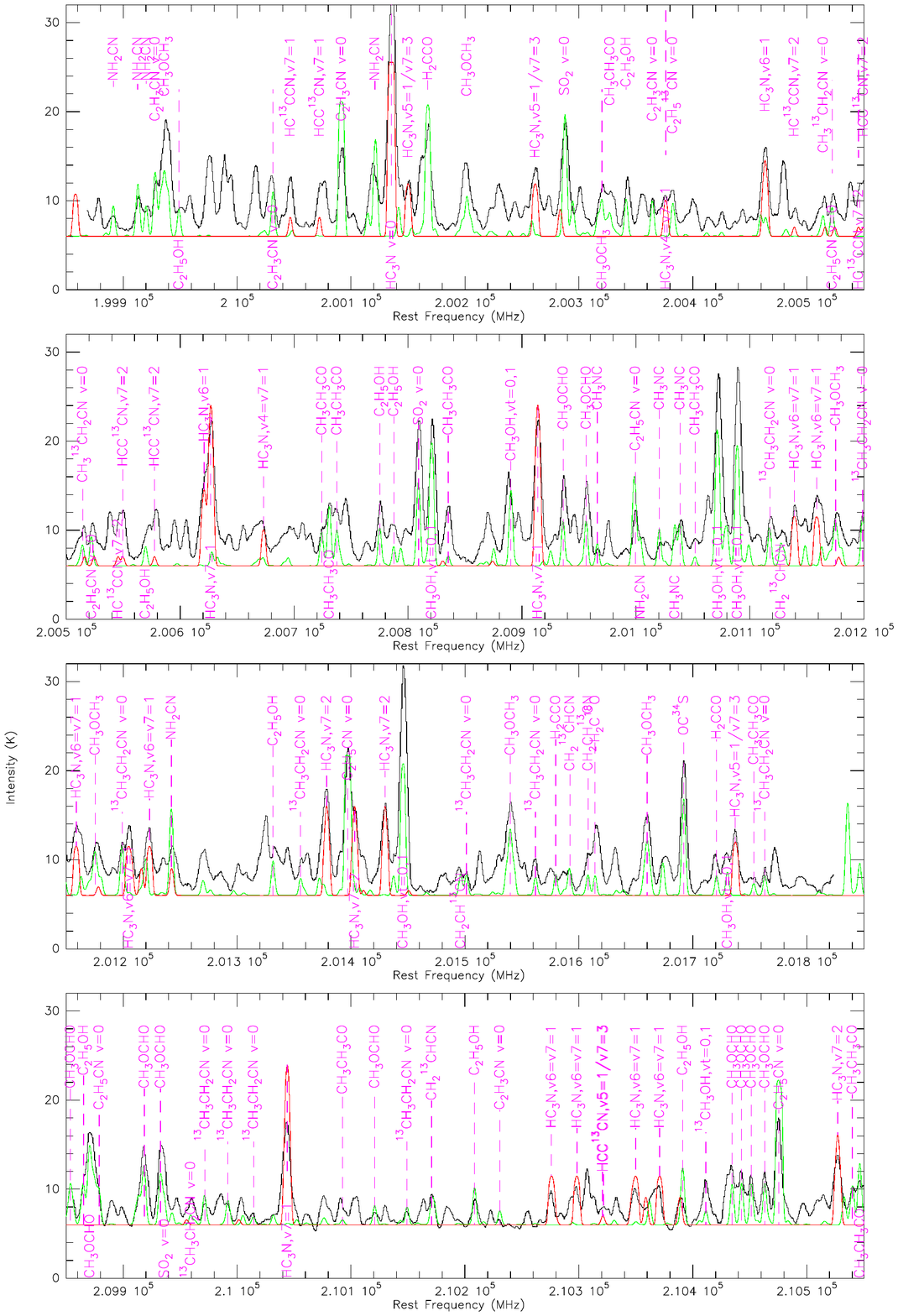} 
 \caption{Full spectral range  towards G10.47+0.03 convolved to $5''$ resolution. The green overlay is the  myXLASS model for all molecules except HC$_3$N, whose emission is overlaid in red.  The displayed panels cover 199.85--201.85 GHz (upper three) and 209.85--210.55 GHz (bottom); the rest is shown in Figs.~\ref{fig:spec2} and \ref{fig:spec3}. }
%  \includegraphics[angle=0,width=0.95\textwidth]{spec1+hc3n.pdf} 
% \caption{201/211 GHz spectral range convolved to $5''$ resolution. The green overlay is the model for all molecules except HC$_3$N, whose emission is overlaid in red.}
  \label{fig:spec1}
\end{figure*}

\begin{figure*}
  \centering
  \includegraphics[angle=0,width=0.99\textwidth]{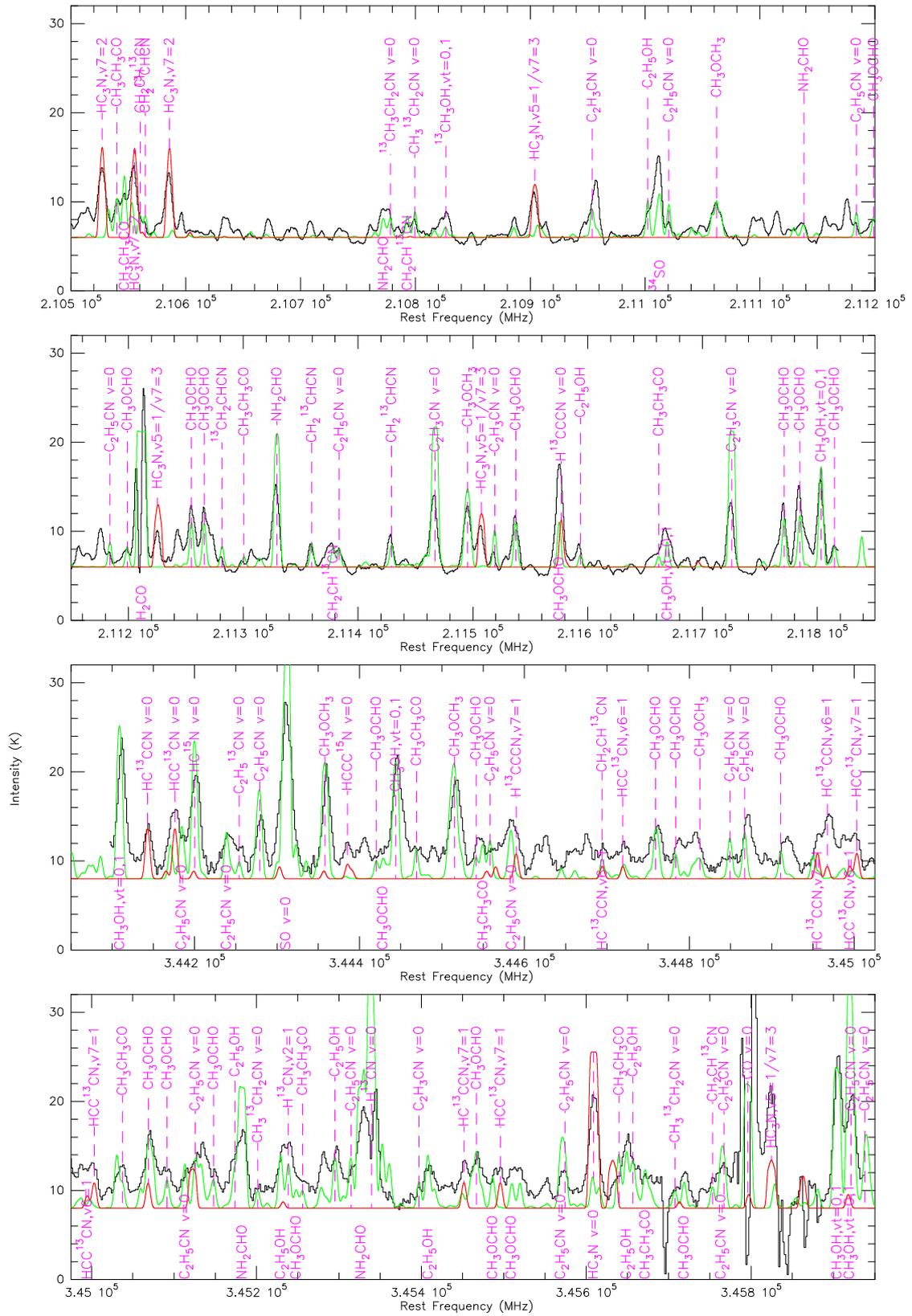} 
 \caption{As Fig.~~\ref{fig:spec1}, displaying 210.5--211.85 GHz (upper two) and 344.05--345.95 GHz (lower two).}
  \label{fig:spec2}
\end{figure*}

\begin{figure*}
  \centering
  \includegraphics[angle=0,width=0.99\textwidth]{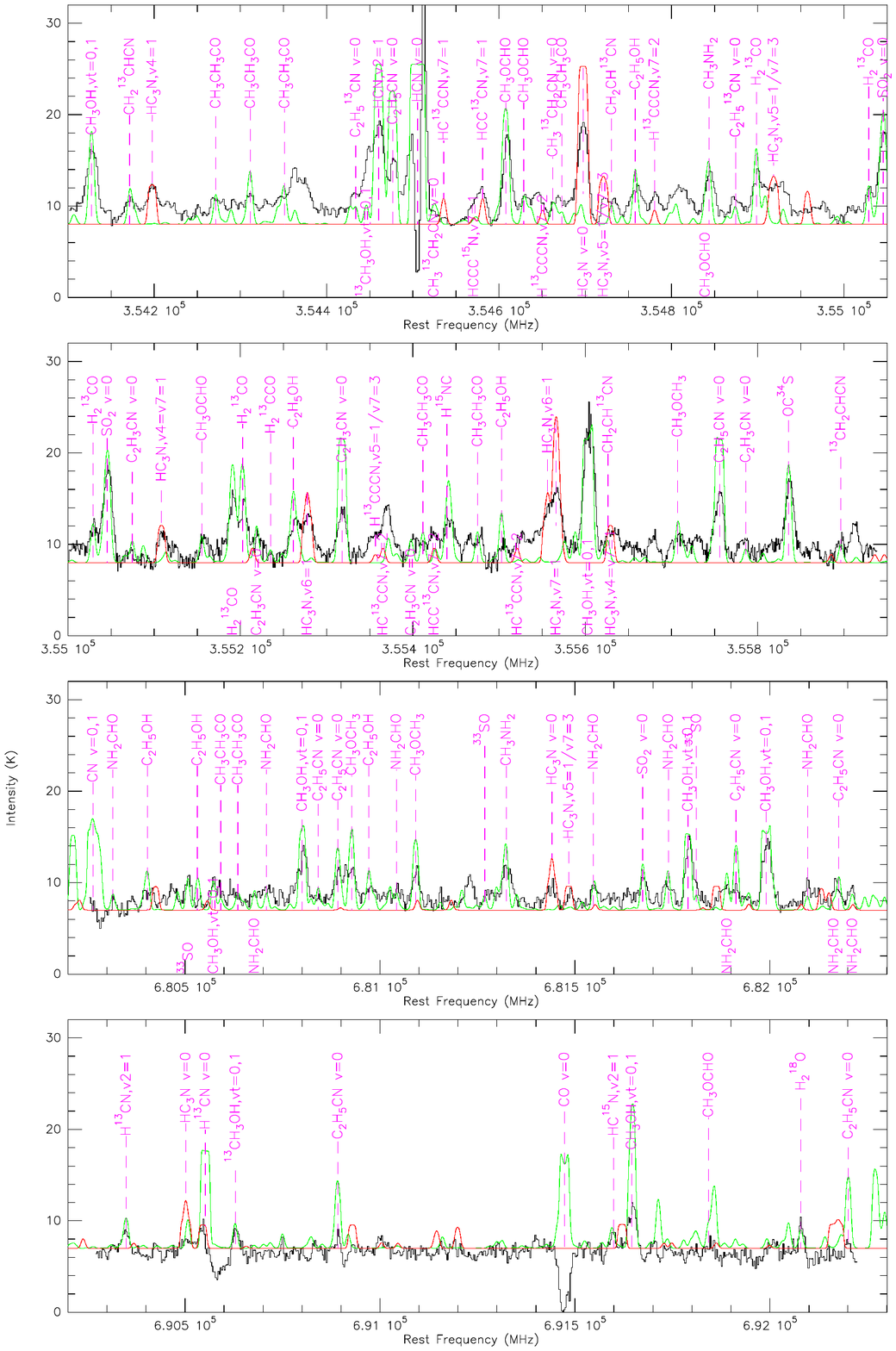} 
 \caption{ As Fig.~~\ref{fig:spec1}, displaying 354.1--355.95 GHz (upper two), 680.2--682.3 GHz (third panel from top), and 690.2--692.3 GHz (bottom).}
  \label{fig:spec3}
\end{figure*}

 In Figs.~\ref{fig:spec1}, \ref{fig:spec2}, and \ref{fig:spec3}, we show the full observed spectral range, convolved to a resolution of 5$''$, and label the identified lines with the molecule (see also Sect.~\ref{sec:ident}).

\section{Line maps}

\begin{figure*}
  \centering
  \includegraphics[bb=63         93        544        749,angle=0,width=0.95\textwidth]{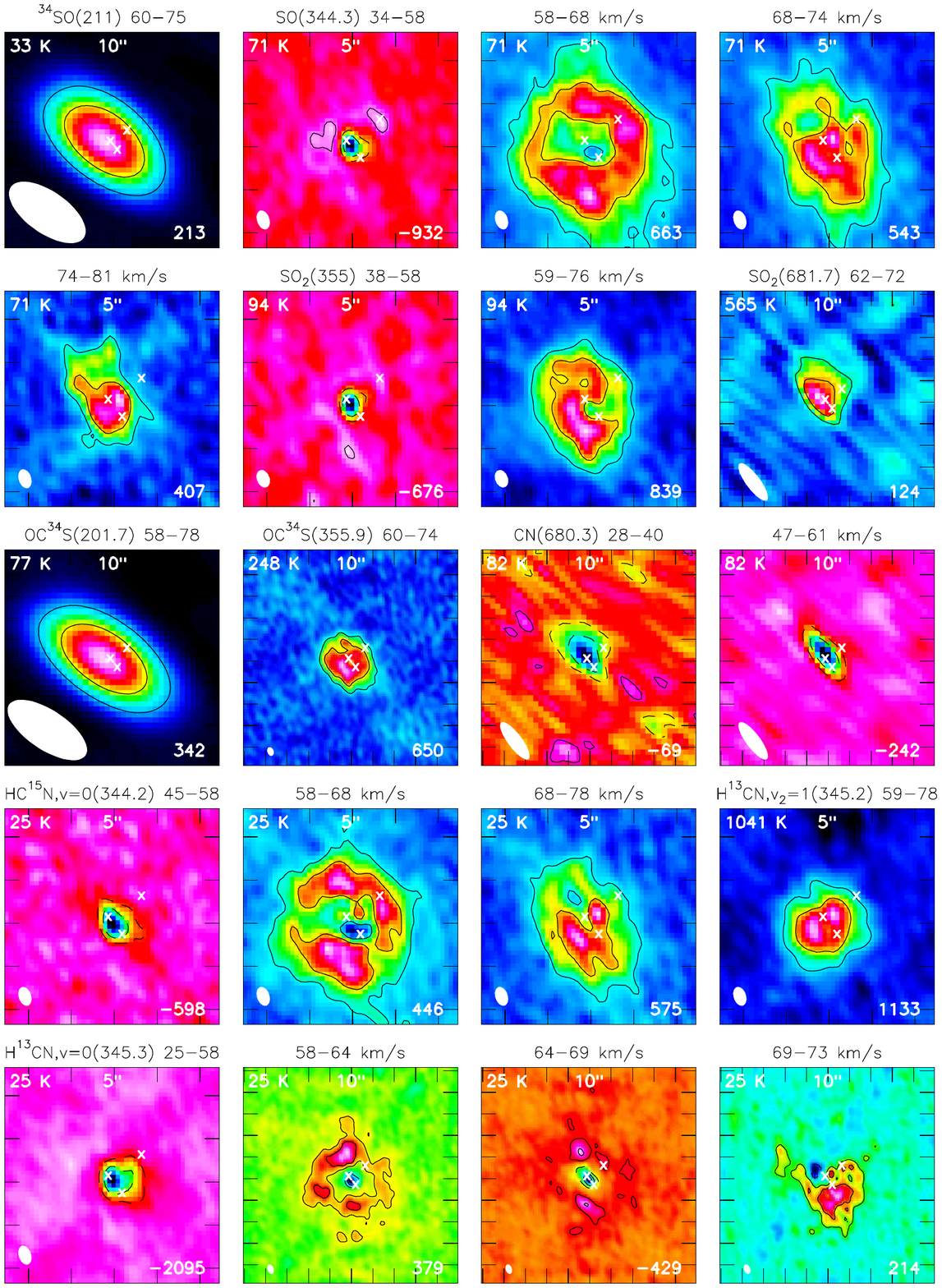} 
 \caption{Integrated line maps.  Above each panel, the molecule, the frequency of the transition (in GHz), and the velocity range (in km~s$^{-1}$) are given. The map size is either 5, 10, or 15$''$, as written in the
   map (tick spaces are $1''$, centered on R.A. 18:08:38.236,
   Dec. -19:51:50.25). Beams are shown in the lower left, and the
   number in the lower right of each panel is the maximum flux in
   K~km~s$^{-1}$ (the contours are $\pm$20 and 50\% of that
   value). The color scale ranges from the minimum to the maximum value.
 The energy of the lower level is given in the upper left. }
  \label{fig:maps1}
\end{figure*}

\begin{figure*}
  \centering
   \includegraphics[bb=63         93        544        749,angle=0,width=0.95\textwidth]{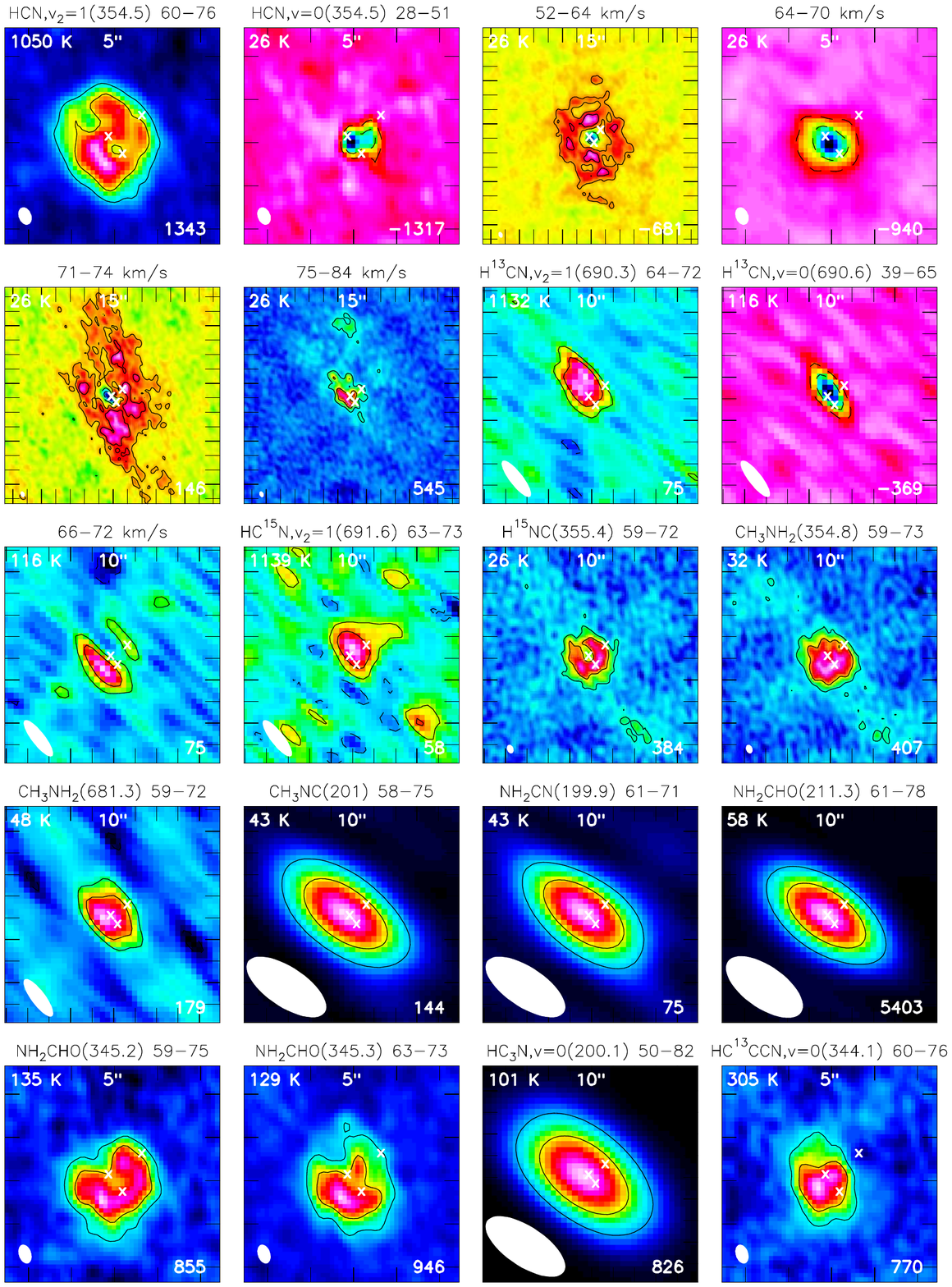} 
 \caption{As Fig.~\ref{fig:maps1}, contd. }
  \label{fig:maps2}
\end{figure*}

\begin{figure*}
  \centering
   \includegraphics[bb=63         93        544        749,angle=0,width=0.95\textwidth]{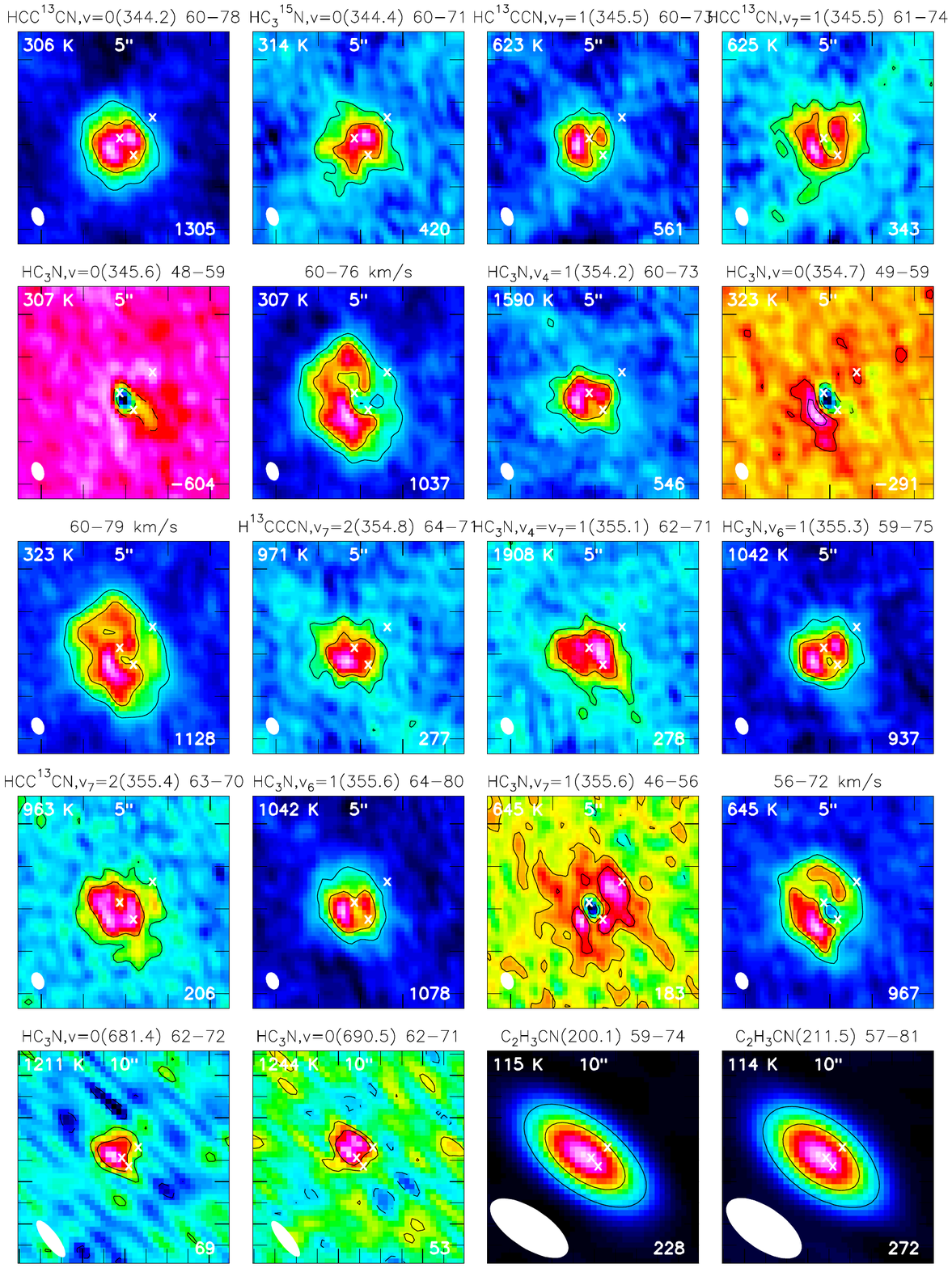} 
 \caption{As Fig.~\ref{fig:maps1}, contd.  }
  \label{fig:maps3}
\end{figure*}

\begin{figure*}
  \centering
   \includegraphics[bb=63         93        544        749,angle=0,width=0.95\textwidth]{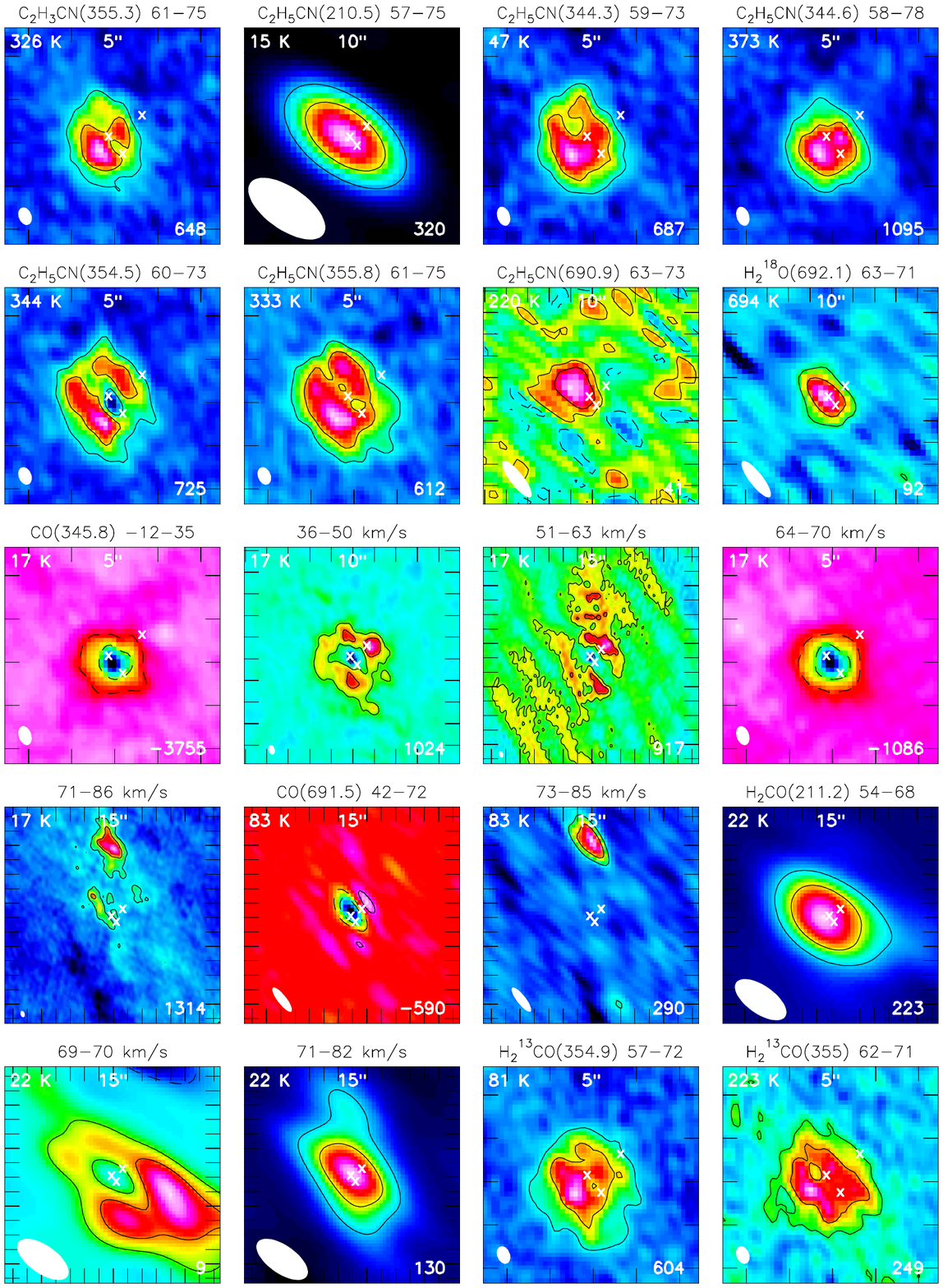} 
 \caption{As Fig.~\ref{fig:maps1}, contd.  }
  \label{fig:maps4}
\end{figure*}

\begin{figure*}
  \centering
   \includegraphics[bb=63         93        544        749,angle=0,width=0.95\textwidth]{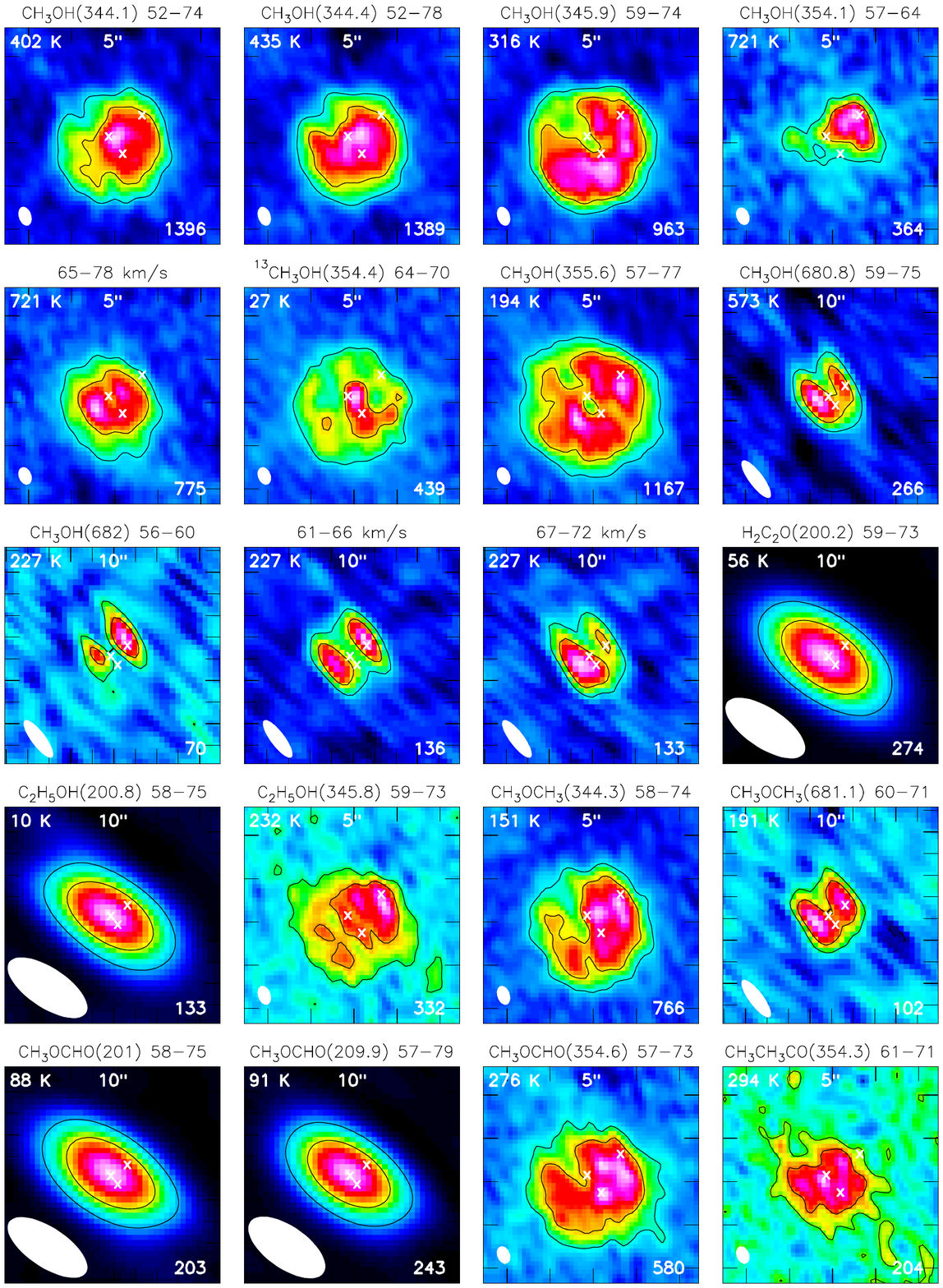} 
 \caption{As Fig.~\ref{fig:maps1}, contd.  }
  \label{fig:maps5}
\end{figure*}

In Figs.~\ref{fig:maps1}, \ref{fig:maps2}, \ref{fig:maps3},  \ref{fig:maps4}, and \ref{fig:maps5} we show 100 integrated line maps, ordered by molecule as in Table~\ref{tab:molecules} and by frequency of the transition (see also Sect.~\ref{sec:linemaps}).

\end{appendix}

\end{document}